\documentclass[aps,prb,twocolumn,superscriptaddress]{revtex4}
\pdfoutput=1
\usepackage[latin9]{inputenc}
\usepackage{graphicx,epsf}
\usepackage{amssymb}
\usepackage{amsmath}
\usepackage{color}
\usepackage{bm}
\usepackage[caption=false]{subfig}
\usepackage{placeins}

\newcommand{\qph}{\quad \phantom{.}}

\newcommand{\w}{\omega}
\newcommand{\onebm}{SBM1}
\newcommand{\twobm}{SBM2}

\newcommand{\bath}{{\rm bath}}
\newcommand{\cpl}{{\rm cpl}}

\newcommand{\HM}{\hat{\mathcal{H}}}
\newcommand{\eff}{{\rm eff}}
\newcommand{\hzc}{h_{z,c}}

\RequirePackage[
   hyperindex,colorlinks,bookmarksnumbered,
   plainpages=true, pdfstartview=FitH,bookmarks=false
]{hyperref}
\hypersetup{linkcolor=blue,urlcolor=blue,citecolor=blue}

\definecolor{darkgreen}{rgb}{0,0.5,0}
\definecolor{darkblue}{rgb}{0,0,0.5}
\definecolor{purple}{rgb}{0.35,0,0.35}
\definecolor{orange}{rgb}{1,0.5,0}
\definecolor{wbcolor}{rgb}{0,.6,1}
\definecolor{todocolor}{rgb}{1,0,0}
\definecolor{jvdcolor}{rgb}{0,0,1}

\newcommand{\Eq}[1]{Eq.~(\ref{#1})}
\newcommand{\Eqs}[1]{Eqs.~(\ref{#1})}

\newcommand{\Sec}[1]{Sec.~\ref{#1}}

\newcommand{\Fig}[1]{Fig.~\ref{#1}}

\newcommand{\Figs}[1]{Figs.~\ref{#1}}

\newcommand{\pdag}{{\phantom{\dagger}}}

\begin{document}
\title{
Two-bath spin-boson model: Phase diagram and critical properties}

\author{Benedikt Bruognolo}
\author{Andreas Weichselbaum}
\author{Cheng Guo}
\author{Jan von Delft}
\affiliation{Physics Department, Arnold Sommerfeld Center for Theoretical Physics, and Center for NanoScience,
Ludwig-Maximilians-Universit\"at, Theresienstra{\ss}e 37, 80333 M\"unchen, Germany}
\author{Imke Schneider}
\affiliation{Physics Department and Research Center OPTIMAS, Technische Universit\"at Kaiserslautern, 67663 Kaiserslautern, Germany}
\author{Matthias Vojta}
\affiliation{Institut f\"ur Theoretische Physik, Technische Universit\"at Dresden, 01062 Dresden, Germany}

\date{\today}

\begin{abstract}
The spin-boson model, describing a two-level system coupled to a bath of harmonic oscillators, is a generic model for quantum dissipation, with manifold applications. It has also been studied as a simple example for an impurity quantum phase transition.
Here we present a detailed study of a U(1)-symmetric two-bath spin-boson model, where two different components of an SU(2) spin 1/2 are coupled to separate dissipative baths. Non-trivial physics arises from the competition of the two dissipation channels, resulting in a variety of phases and quantum phase transitions. We employ a combination of analytical and numerical techniques to determine the properties of both the stable phases and the quantum critical points.
In particular, we find a critical intermediate-coupling phase which is bounded by a
continuous quantum phase transition which violates the quantum-to-classical
correspondence.
\end{abstract}


\maketitle


\section{Introduction}

Impurity models, describing small quantum systems coupled to one or multiple baths of bosons or fermions, have seen a lot of activity over the last years, for a variety of reasons:
(i) Impurity models display a rich phenomenology, including local Fermi-liquid and non-Fermi-liquid behavior,\cite{hewson,NB} phase transitions and quantum criticality,\cite{mvrev,logan14} as well as interesting properties far from equilibrium.\cite{rosch03}
(ii) Impurity models can often be simulated by numerical means more efficiently than lattice models,\cite{wilson_rev_1975,bulla_rev_2008} such that, on the one hand, high-accuracy numerical results can guide analytical approaches and, on the other hand, analytical concepts can be readily tested numerically. A particularly interesting branch is non-equilibrium physics where quantum impurity models have served a test bed for methodological developments.
(iii) Impurity models find realizations in diverse settings such as dilute magnetic moments in bulk solids,\cite{haas30,rosch08} electrons in quantum dots coupled to leads,\cite{cronen98,gg98} quantum bits in a dissipative environment,\cite{makhlin} and charge-transfer processes in organic molecules.\cite{garg85} The design of impurity models in cold-atom systems provides further means of manipulating and detecting impurity phenomena.\cite{demler13,nishida13}

The spin-boson model (SBM1 in the following) is a simple paradigmatic model for quantum dissipative systems.\cite{leggett} It describes a two-level system, i.e., a spin 1/2, which is coupled to both a bath of harmonic oscillators and a transverse field. While the field induces tunneling (i.e.\ delocalization) between the two states, the oscillator bath causes friction and impedes tunneling. For gapless baths, characterized by a power-law spectral density $J(\w)\propto\w^s$ with $0<s\leq 1$, this competition results in a quantum phase transition between a delocalized and a localized phase which has been studied extensively.\cite{kehrein_spin-boson_1996,BTV03,VTB,VTB_err,fehske,winter09,guo_vmps_2012,flow,vojta_NRG_2012} As has been shown both analytically and numerically,\cite{fehske,winter09,VTB_err,flow,vojta_NRG_2012} this quantum phase transition obeys the so-called quantum-to-classical correspondence: It is equivalent to the thermal phase transition of a classical Ising chain with long-ranged interactions falling off as $1/r^{1+s}$ where $r$ is the distance between two classical spins.\cite{fisher_critical_1972,koster,luijten_classical_1997}

In this paper we consider the generalization of the spin-boson model to two baths ($i=x,y$ below),\cite{anirvan,rg_bfk,antonio_xy} dubbed {\twobm}. It is described by
$\HM = \HM_{\rm s} + \HM_{\cpl} + \HM_{\bath}$ with
\begin{subequations}
\label{eq:h}
\begin{eqnarray}
\HM_{\rm s} &=& - \vec{h}\cdot \frac{\vec{\sigma}}{2}\,,\\
\HM_{\cpl} &=& \sum_{i=x,y} \sum_{q}  \lambda_{q i} \frac{\sigma_{i}}{2}
( \hat{a}^\pdag_{q i} + \hat{a}_{q i}^{\dagger} )\,, \\
\HM_{\bath}  &=&
   \sum_{i = x,y} \sum_{q}
\w_q \hat{a}_{q i}^{\dagger} \hat{a}_{q i}^\pdag \,.
\qph
\end{eqnarray}
\end{subequations}
The two-level system (or quantum spin, with $\sigma_{x,y,z}$ being the
vector of Pauli matrices) is coupled both to an external field
$\vec{h}$ and, via $\sigma_x$ and $\sigma_y$, to two independent
bosonic baths, whose spectral densities $J_i (\omega) = \pi \sum_{q}
\lambda_{q i}^{2} \delta(\w -\w_q)$ are assumed to be of the same power-law
form,
\begin{equation}
J_i(\w) = 2\pi\, \alpha_i\, \omega_{\rm c}^{1-s} \, \omega^s\,, \quad
 0<\w<\w_{\rm c}\,,
\label{power}
\end{equation}
where $\w_{\rm c}=1$ defines the unit of energy used throughout the paper.
For a symmetric coupling to identical bath, i.e.\,$\alpha=\alpha_x=\alpha_y$, and $h_x=h_y=0$ the model displays a U(1) symmetry, corresponding to a rotation of the impurity spin about its $z$ axis combined with corresponding bath-mode rotation.
In addition, the model features a separate Z$_2$ symmetry for $h_z=0$, corresponding to $\sigma_z \leftrightarrow -\sigma_z$.

The model {\twobm} is governed not only by the competition between the local field, which tends to point the spin in the $\vec{h}$ direction, and the dissipative bath effects, but also by a competition between the two baths, as an oscillator bath which couples to $\sigma_i$ tends to localize the spin in $i$ direction. As a result, the combined dissipative effect of both baths in {\twobm} can be smaller than that of one bath alone (in a sense which will become clear in the course of the paper) -- an effect which has been dubbed ``frustration of decoherence''.\cite{antonio_xy}
In practical realizations of {\twobm}, the two baths can be two different sources of dissipation influencing a quantum bit,\cite{antonio_xy,dima} or two spin-wave modes which couple to a magnetic impurity in a magnet.\cite{sushkov98,VBS}

The model {\twobm} is of particular theoretical interest because it displays a non-trivial intermediate-coupling (i.e. critical) phase, characterized by partial screening of the impurity degree of freedom corresponding to a fractional residual moment, not unlike in the two-channel Kondo state.\cite{NB,andrei1984,wiegmann1985}
The existence of this critical phase, originally deduced by perturbative RG arguments,\cite{anirvan,VBS,rg_bfk} was recently confirmed numerically.\cite{guo_vmps_2012} The latter study, performed using a variational matrix-product-state (VMPS) approach, also revealed that the critical phase is unstable at large couplings, resulting in a complex phase diagram.

It is the purpose of this paper to study the physics of {\twobm} in some detail, extending the results published in Ref.~\onlinecite{guo_vmps_2012}, with particular focus on the quantum phase transitions occurring in this model. To this end, we combine VMPS calculations with analytical renormalization-group and scaling approaches. Our implementation of VMPS, including the use of the U(1) symmetry and an optimized boson basis, enables highly accurate studies of quantum critical behavior.


\subsection{Summary of results}

\begin{figure}[h!]
\centering
\includegraphics[width=.5\textwidth]{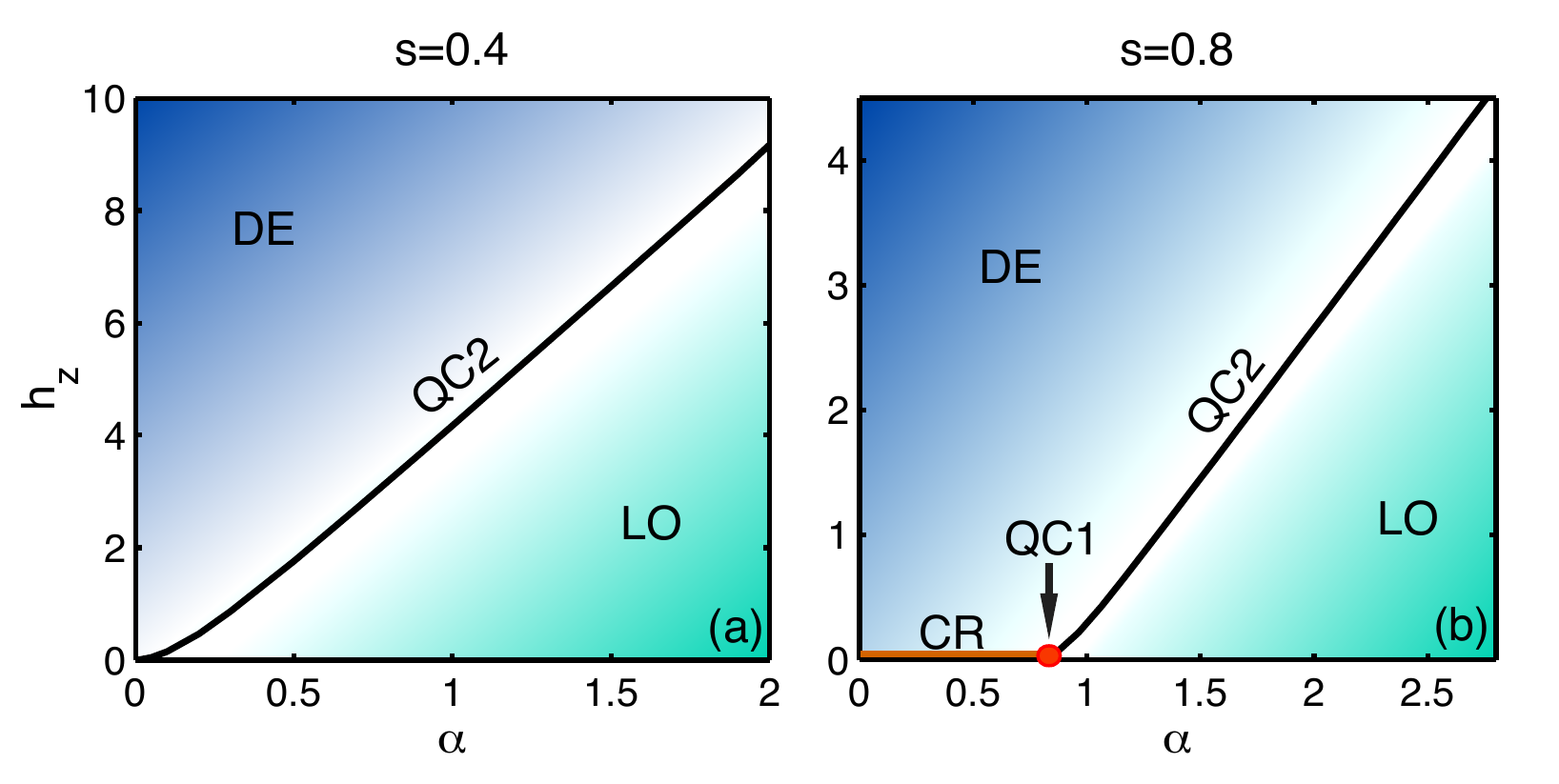}
\caption[]{
Quantitative phase diagrams of {\twobm} for $s=0.4$ (a) and $s=0.8$ (b). For a bath exponent $s<s^*\approx0.76$ in (a), the {\twobm} ground state is either in a delocalized (DE) or localized (LO) phase depending on coupling strength $\alpha$ and magnetic field $h_z$; the corresponding quantum phase transition is controlled by the critical fixed point QC2. For larger $s>s^*$ in (b) an additional critical phase (CR) emerges at $h_z=0$ and small couplings. The quantum phase transition between LO and CR is controlled by a different critical fixed point QC1.
}
\label{fig:PhaseDiag}
\end{figure}

We have used VMPS to determine quantitative phase diagrams for the U(1)-symmetric version of SBM2 as function of the bath exponent $s$, the dissipation strength $\alpha$, and the transverse field $h_z$.
For $0<s<1$ and finite $h_z$, there is always a transition between a delocalized (DE) and a localized (LO) phase, Fig.~\ref{fig:PhaseDiag} with the LO phase spontaneously breaking the model's U(1) symmetry.
There is no localization for $s=1$ (not shown)\cite{antonio_xy} -- this is qualitatively different from the behavior of the standard single-bath spin-boson model (SBM1) and reflects the frustration of decoherence mentioned above.
For $h_z=0$ the critical (CR) phase emerges, existing for $s^\ast<s<1$ and small $\alpha$.

Based on numerical and analytical results for the quantum critical behavior, we conclude that the transition between the DE and LO phases, controlled by a fixed point labelled QC2 in the body of the paper, is in the universality class of the classical XY chain with $1/r^{1+s}$ interactions, i.e., obeys a quantum-to-classical correspondence. In particular, $s=1/2$ corresponds to the upper-critical dimension for this transition, with mean-field behavior found for $s<1/2$.

In contrast, the transition between CR and LO, controlled by a different fixed point QC1, does not appear to obey a quantum-to-classical correspondence. Its exponents fulfill hyperscaling relations for $h_z=0$, but hyperscaling is violated in the presence of a transverse field. We propose how to construct a critical field theory which should ultimately enable an analytical understanding of this conceptually interesting non-classical transition.


\subsection{Outline}

The body of the paper is organized as follows.
In Section~\ref{sec:method} we introduce the employed VMPS method. In particular we discuss both the variational choice of bosonic basis states and the implementation of the U(1) symmetry into the algorithm.
Section~\ref{sec:phases} describes the phase diagram of the U(1)-symmetric {\twobm}, together with the main characteristics of the stable phases.
The subsequent Section~\ref{sec:flow} analyzes the numerical findings in terms of renormalization-group flow and discusses the resulting quantum critical points.
Section~\ref{sec:RG} is devoted to analytical approaches to the critical phenomena of {\twobm}, using the toolbox of field theory and epsilon expansion. In particular, we highlight that QC2 is expected to follow the quantum-to-classical correspondence while QC1 is not.
In Section~\ref{sec:numerical} we show numerical results for critical properties of {\twobm}. We will extract numerous critical exponents as function of the bath exponent $s$, confirming the analytical expectations.
The concluding Section~\ref{sec:concl} will highlight open problems as well as connections to other impurity and lattice problems. In addition, the physics of {\twobm} with broken U(1) symmetry will be quickly discussed. Technical details are relegated to various appendices.


\section{VMPS method}
\label{sec:method}

We start by describing the numerical VMPS approach which we employed to study \twobm. This extends the corresponding presentation in Ref.~\onlinecite{guo_vmps_2012supp}. In particular the explicit implementation of the U(1) symmetry, which we found crucial to obtain accurate critical exponents, is a novel ingredient here.

\subsection{Discretization and Wilson chain mapping}

Since both bosonic baths of {\twobm} are non-interacting and gapless, it is possible to transfer the concept of energy-scale separation frequently employed in Numerical Renormalization Group (NRG).\cite{wilson_rev_1975,bulla_rev_2008} To this end, the spectral functions of the baths  are logarithmically discretized. Then the Hamiltonian is mapped on a semi-infinite tight binding chain, a so-called Wilson chain.

The choice of a logarithmic coarse graining of the spectral function $J_i$ is motivated by the fact that the study of critical behavior requires exponentially small energy scales. To resolve these scales appropriately, a logarithmic coarse-graining is necessary, since it yields an exponentially enhanced low-energy resolution compared to a linear or power-law discretization.
Assuming the spectral function $J_i$ of each bosonic bath has a non-zero contribution for energies $\omega \in ]0,\omega_c]$, with $\omega_c=1$ being an upper cut-off frequency, we introduce a dimensionless discretization parameter $\Lambda>1$ which defines a set of intervals with discretization points,\cite{wilson_rev_1975,bulla_rev_2008,BTV03,BLTV05}
\begin{eqnarray}
\omega_0^z =& \omega_c \quad &(m=0)\,, \nonumber \\
\omega_m^z =& \omega_c\Lambda^{-m+z} \quad &(m=1,2,3,...)\,,
\end{eqnarray}
with $z \in [0,1[$ an arbitrary shift. Averaging over different $z$ uniformly distributed in $[0,1[$ is referred to as $z$-averaging\cite{}. Considering a symmetric coupling of the impurity to two identical baths and using $z=0$ for simplicity, the discretized Hamiltonian is represented by
\begin{eqnarray}
\label{eq:hstar}
\HM_\bath
&=&
   \sum_{i = x,y} \sum^{\infty}_{m=0} \left[
\xi_m \hat{a}_{m i}^{\dagger} \hat{a}_{m i}^\pdag +
\gamma_{m} \frac{\sigma_{i}}{2}
( \hat{a}^\pdag_{m i} + \hat{a}_{m i}^{\dagger} )\right]\,,
\qph
\end{eqnarray}
with $\hat{a}_{m i}$ being a discrete bosonic state at energy $\xi_m$ and coupling strength $\gamma_m$ to the impurity spin. For general $J(\omega)$ one has\cite{BLTV05}
\begin{subequations}
\begin{eqnarray}
\gamma_m^2 &=& \int_{\omega_{m+1}}^{\omega_{m}} J(\omega) d\omega\,,  \\
\xi_m &=&\gamma_m^{-2} \int_{\omega_{m+1}}^{\omega_{m}} \omega J(\omega)  d\omega\,.
\end{eqnarray}
\end{subequations}
Employing the improved $z$-averaging scheme of \v{Z}itko and Pruschke to reduce discretization artifacts,\cite{zitko_energy_2009} the explicit expressions for the parameters for general $z$ are given by\cite{guo_vmps_2012supp}
\begin{subequations}
\begin{eqnarray}
\xi^z_0 =& \Big[\tfrac{1-\Lambda^{z(1+s)}}{(1+s)\ln\Lambda} - z +1  \Big]^{\frac{1}{1+s}} \quad &(m=0)\,,	\nonumber \\
\xi^z_m =& \Big[\tfrac{\Lambda^{-(s+1)(m+z)}( \Lambda^{(1+s)}-1 )}{(1+s) \ln\Lambda}  \Big]^{\frac{1}{1+s}} \sim \omega^z_m \quad &(m>0)\,, \nonumber \\
\end{eqnarray}
\begin{eqnarray}
\gamma_0^z =& \sqrt{\tfrac{2\pi\alpha}{1+s} ( 1-\Lambda^{-z(1+s)} )} \quad &(m=0)\,, \nonumber\\	
\gamma_m^z =& \sqrt{\tfrac{2\pi\alpha}{1+s}(\Lambda^{1+s}-1)\Lambda^{-(m+z)(1+s)}} \sim (\omega^z_m)^{\frac{s+1}{s}} &(m>0)\,. \nonumber \\
\end{eqnarray}
\end{subequations}
Following the standard NRG protocol, the discretized Hamiltonian in Eq.\,(\ref{eq:hstar}) is mapped using an exact unitary transformation onto a semi-infinite tight-binding chain, dubbed Wilson chain, with the impurity coupled to the open end only. The resulting Hamiltonian including $(N+1)$ bosonic sites is given by $\HM_N\cong\HM_s + \HM_{\cpl}+\HM_{\bath}^{(N)}$ with
\begin{subequations}
\label{eq:hchain}
\begin{eqnarray}
 \label{eq:hchaincpl}
\HM_{\cpl}&=& \sum_{i=x,y} \sqrt{\frac{\eta_0}{\pi}} \frac{\sigma_i}{2} (\hat{b}_{0 i}+\hat{b}^{\dagger}_{0 i} )\,,  \\
\HM^{(N)}_{\bath}&=&  \sum_{i=x,y} \Big[ \sum^{N}_{k=0} \epsilon_{k} \hat{n}_{ki}  +  \sum^{N-1}_{k=0} (t_{k}\hat{b}^{\dagger}_{ki}\hat{b}_{(k+1) i} + \text{H.c.}) \Big]\,, \nonumber \\  \label{eq:hchainbath}
\end{eqnarray}
\end{subequations}
with the operator $\hat{n}_{ki} = \hat{b}_{ki}^{\dagger}\hat{b}_{ki}$ counting the number of bosons of bath $i$ on chain site $k$. Each bosonic site represents a harmonic oscillator at frequency $\epsilon_k \sim \Lambda^{-k}$ that is coupled to its nearest neighbors by the hopping amplitude $t_k\sim \Lambda^{-k}$. Assuming identical baths, $\eta_0 = \int J(\omega)d\omega$ describes the overall coupling between a bath and impurity. Note that the impurity spin now couples to a single bosonic degree of freedom located at $k=0$, i.e.\,the first site of a bosonic tight-binding chain (see also \Fig{fig:VMPSfig2} below). Their combined local Hamiltonian is given by $\HM_0$.


\subsection{VMPS optimization with OBB}

The steps remaining in the NRG procedure would involve an iterative diagonalization by adding one site at a time and a subsequent truncation of the high-energy states of the system, keeping only the $D$ lowest lying energy eigenstates. However, the bosonic nature of the model complicates the NRG approach drastically. Employing NRG, it is required to truncate the infinite dimensional local bosonic Hilbert spaces on site $k$ to manageable number of $d_k$ states. Thus, a priori, NRG is not able to take into account the growing oscillator displacements $\hat{x}_{k i} = 1/\sqrt{2}(\hat{b}_{k i} + \hat{b}_{k i})$ occurring in the system's localized phase. This restricts its application to the delocalized phase. Already at the phase boundary, in combination with the inherent mass-flow error\cite{flow}, this leads to non-mean-field results for the critical exponents of SBM1 in the regime $s<1/2$.\cite{VTB_err,vojta_NRG_2012}

To resolve the issue of bosonic state space truncation, Guo et al.\cite{guo_vmps_2012} proposed a variational matrix-product-state (VMPS) approach involving an optimized boson basis (OBB), that allows an accurate numerical study of the entire phase diagram in the (generalized) spin-boson model. Since we heavily used this method for the numerical results presented here for \twobm, we briefly outline the concept of this powerful approach.\cite{guo_vmps_2012supp}

\begin{figure}[h!]
\centering
\includegraphics[width=0.45\textwidth]{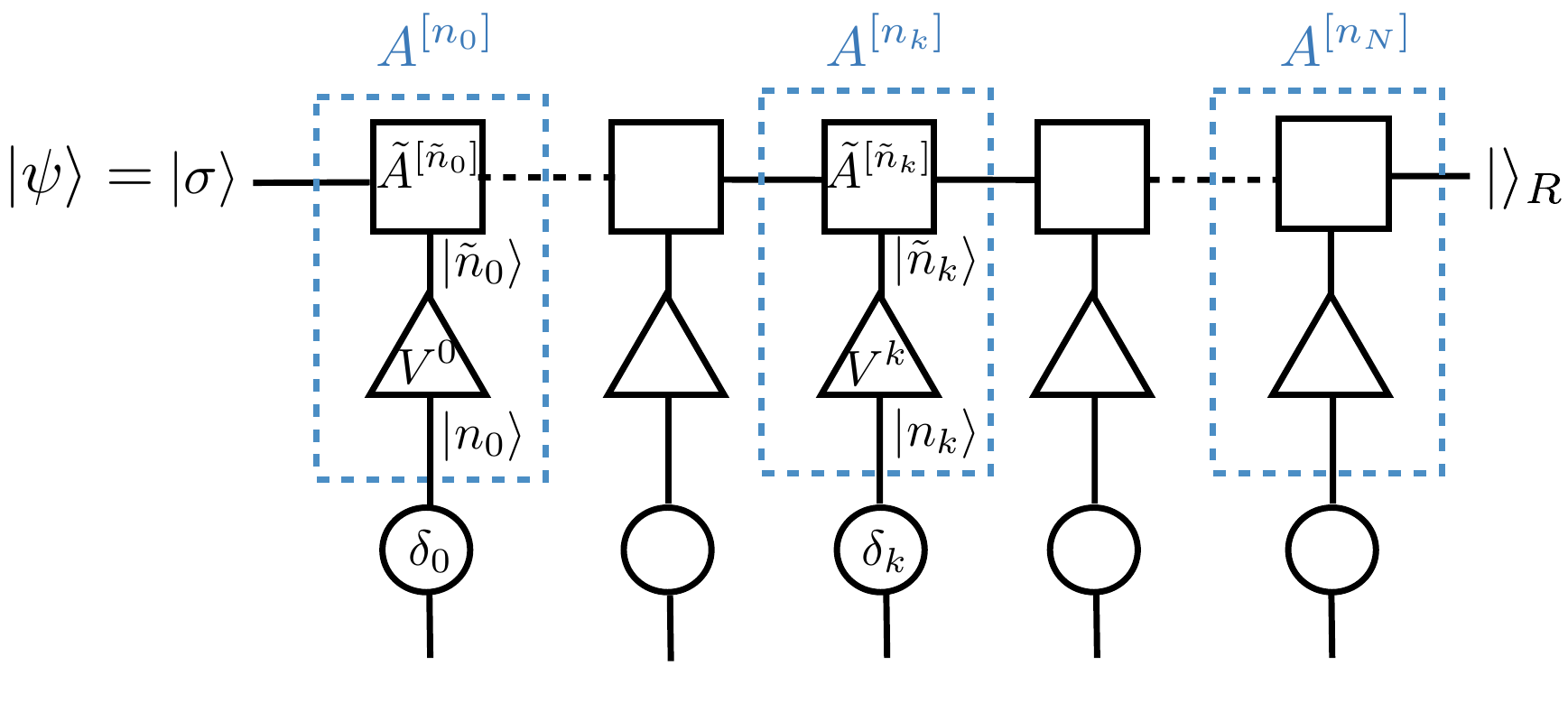}
\vspace{-10pt}
\caption[]{Schematic diagram of $|\psi\rangle$ in \Eq{eq:MPS1} using the OBB representation with explicit bosonic shift. The first index of $A^{[n_0]}$ and the last index of $A^{[n_N]}$ link to the impurity and the right-vacuum state $| \rangle_R$, respectively. }
\label{fig:VMPSfig2}
\end{figure}

The starting point of the variational procedure is setting up an initially random many-body state $|\psi\rangle$ of the truncated Wilson chain described by $\HM_N$ [having $(N+1)$ sites in total] in the language of matrix-product states (MPS):\cite{schollwock_2011}
\begin{equation}
\label{eq:MPS1}
|\psi\rangle = \sum_{\sigma=\uparrow,\downarrow} \sum_{n} \Big(A^{[n_0]} A^{[n_1]} ... A^{[n_{N}]} \Big)_{\sigma} |\sigma\rangle |n\rangle\,,
\end{equation}
where $|\sigma\rangle = |$$\uparrow$$\rangle, |$$\downarrow$$\rangle$ are the eigenstates of $\sigma_x$ and the states $|n\rangle = |n_0,...,n_{N}\rangle$ represent the boson-number eigenstates of the truncated Fock space, i.e. $\hat{n}_{ki} |n\rangle=n_{ki} |n\rangle$ with $n_{ki} = 0,...,d_{k} - 1$. Combining the state spaces of both chains in Eq.~\eqref{eq:hchaincpl} and \eqref{eq:hchainbath} to supersites, $n_k=(n_{k x},n_{k y})$ should be interpreted as a combined index of the x- and y-chain. Each $A^{[n_k]}$ forms a $D\times D$ matrix with elements $(A^{[n_k]})_{\alpha \beta}$, except for $A^{[n_0]}$ and $A^{[n_{N}]}$ connecting to local impurity and vacuum states respectively, as indicated in \Fig{fig:VMPSfig2}. Using standard MPS methods, we optimize $|\psi\rangle$ by iteratively varying one $A^{[n_k]}$ at a time in order to find an appropriate representation of the ground state of $\HM_N$.

The main advantage of VMPS is the possibility to change the local basis during the step-wise optimization process, while NRG in principle requires the local basis to be fixed a priori before starting to diagonalize. To exploit this, we implement the OBB using two key features:
\\

\noindent{1}.\ \emph{Effective local basis}: a basis transformation $V$ is introduced with $V^{\dagger}V=\mathbb{I}$, which maps the local harmonic oscillator basis $|n_k\rangle$ onto a smaller effective basis $|\tilde{n}_k\rangle$ on each site $k$,
\begin{equation}
|\tilde{n}_k\rangle = \sum_{n_k=0}^{d_k-1} V_{\tilde{n}_k,n_k} |n_k\rangle \quad (\tilde{n}_k =0,\,...\,,\tilde{d}_k-1)\,, \\
\end{equation}
with $d_k$ and $\tilde{d}_{k}$ denoting the size of the original and effective basis, respectively. Merging $V$ into the $A$-tensors on each bosonic site, the structure of $A^{[n_k]}$ in \Eq{eq:MPS1} is then given by
\begin{equation}
A^{[n_k]}_{\alpha,\beta} = \sum_{\tilde{n}_k=0}^{\tilde{d}_{k}-1} \tilde{A}^{[\tilde{n}_k]}_{\alpha,\beta} V_{\tilde{n}_k,n_k}\,.
\label{eq:MPS_OBB}
\end{equation}
Nevertheless, from an efficiency point of view, it is desirable to keep the separate structure of $\tilde{A}$ and $V$, where $\tilde{A}^{[\tilde{n}_k]}$ links the effective bosonic basis to the left and right part of the chain, while $V$ maps the original to the effective local basis. The local optimization procedure for each site thus splits into two steps: at first, $V$ is updated and in this process the optimal effective local basis set $|\tilde{n}_k\rangle$ is determined. Then we optimize $\tilde{A}^{[\tilde{n}_k]}$ using the new local basis states and move to the next site.\\
 Note that with the introduction of the OBB  a second adjustable dimension $\tilde{d}_k$ besides the bond dimension $D$ exists. Treating $\tilde{A}$ and $V$ as separate structures, both dimensions are fixed before the start of the ground-state optimization. If a dynamical adjustment of the bond dimensions is required, one has to switch to a two-side optimization procedure or variants of these, which is numerically more expensive.\cite{schollwock_2011} This is for example necessary when enforcing explicit symmetry conservation.  In practice, this implementation makes an increase of the size of the local basis sets from $d_k \approx 10$ to $d_k \lesssim 10^{4}$ possible, while using typically $\tilde{d}_k\lesssim d_k$ below. \\

\noindent{2}.\ \emph{Oscillator shifts}: moreover, in the localized phase we incorporate an oscillator shift in the Hamiltonian to take the oscillator displacement  into account. The oscillator coordinates $\hat{x}_{ki}=1/\sqrt{2}(\hat{b}_{ki}+\hat{b}_{ki}^{\dagger})$ are shifted by their equilibrium value  $\langle\hat{x}_{ki}\rangle$\cite{fehske} to be determined self-consistently in a variational setting, such that OBB captures the quantum fluctuations near the shifted coordinate $\hat{x}_{ki}'=\hat{x}_{ki}'-\langle\hat{x}_{ki}\rangle$. This is achieved by formulating the shift $\delta_{ki}$ as unitary transformation acting on the Hamiltonian itself. With
\begin{equation}
\hat{U}(\delta_{ki}) = e^{\frac{\delta_{ki}}{2}(\hat{b}^{\dagger}_{ki}-\hat{b}_{ki})}\,,
\end{equation}
the shifted local bosonic operators $\hat{b}_{ki}'^{\dagger}$ and $\hat{b}_{ki}'$ are
\begin{eqnarray}
\hat{b}_{ki}'&\equiv& \hat{U}^{\dagger}(\delta_{ki}) \hat{b}_{ki} \hat{U}(\delta_{ki}) = \hat{b}_{ki} +\frac{\delta_{ki}}{\sqrt{2}}\,. \label{eq:b}
\end{eqnarray}
By the application of $\hat{U}(\delta_{ki})$ we automatically shift $\hat{x}_{ki}$ by $\delta_{ki}$,
\begin{equation}
 \hat{x}_{ki}' = \frac{1}{\sqrt{2}}(\hat{b}_{ki}'+\hat{b}_{ki}^{\dagger'}) = \hat{x}_{ki} +\delta_{ki}\,.
\end{equation}
 After processing the local optimization procedure, we calculate the mean displacement $\langle\hat{x}_{ki}\rangle$. By setting $\delta_{ki} = -\langle\hat{x}_{ki}\rangle$ and replacing $\hat{b}_{ki}$ with the displaced $\hat{b}'_{ki}$, the shift is included exactly on the Hamiltonian level, $ \hat{U}^{\dagger}(\delta_{ki})\HM_{N}(\{\hat{b}_{ki}\})\hat{U}(\delta_{ki}) = \HM_N (\{\hat{b}'_{ki}\}) = \HM_{N}'(\{\hat{b}_{ki}\},\{\delta_{ki}\})$. Afterwards, the optimization of the current site is repeated in the shifted local bosonic basis until $\langle\hat{x}_{ki}\rangle$ converges, before moving to the next site. \\

\noindent{T}he implementation of an OBB with shifted oscillator modes allows us to simulate an effective local basis that would require a local dimension of $d^{\mathrm{eff}}_k\approx10^{10}$ in the non-shifted basis, while the actual shifted basis can be kept small, $d_k \lesssim 10^{2}$. In addition, since the variational procedure determines the  optimal shift $\delta_{ki}$ for each site of the Wilson chain individually, the exponential growth of $\langle \hat{x}_{ki} \rangle \propto \Lambda^{k}$ with increasing iteration number $k$ no longer represents a barrier for the method.

Working in the Wilson chain setup with an exponentially decreasing energy scale, it is advantageous to replicate the NRG rescaling procedure in the iterative VMPS procedure in order to avoid losing numerical accuracy towards higher iterations. Therefore, when optimizing $A^{[n_k]}$, we rescale the Hamiltonian in the local picture by a factor $\Lambda^k$ to ensure that optimization can take place on the effective energy scale $\sim \omega_c$.

Employing standard VMPS methods, we determine the convergence of $|\psi\rangle$ by calculating the variance of the (unscaled) energy $E^0_k$ of the ground state calculated at each site $k$. The iterative optimization procedure is stopped, once ${\rm std}(E^0_k)/\bar{E}^0<\epsilon$, using double precision accuracy $\epsilon =10^{-15}$ with $N=50,  \Lambda=2$ and thus $\epsilon_{N-1}\sim\Lambda^{-N-1}=10^{-15}$. The resulting state $|\psi\rangle$ is considered to be a reliable approximation of the system's ground state given $\HM_N$. When computing systems where the effective energy resolution drops below double precision, the relevance of numerical noise as a perturbation to $\HM_N$ should be double-checked by additionally studying the energy-flow diagrams.

Most results shown in this paper have been obtained using parameters $\Lambda=2$, $N=50$, $\tilde{d}_k=24$, unless noted otherwise.


\subsection{U(1) symmetry}
\label{sec:U1}

Considering the case with no in-plane magnetic fields, $h_x=h_y=0$, the system exhibits an Abelian U(1) symmetry: The Hamiltonian is invariant under simultaneous rotation of the impurity spin and the bosonic baths in the $xy$-plane by an arbitrary angle $\phi$, leading to a two-fold degeneracy of the resulting ground state. A rotation of this type is described by a unitary operator $\hat{U}(\phi)$,
\begin{equation}
| \psi \rangle \rightarrow \underbrace{e^{i\phi \hat{S}}}_{\equiv \hat{U}(\phi)} |\psi \rangle \label{U1trafo}\,,
\end{equation}
where $\hat{S}$  is the generator of the continuous U(1) symmetry, given by
\begin{equation}
 \label{eq:u1_generator}
\hat{S} = \frac{1}{2} \sigma_z + i \sum_k \big( \hat{b}^{\dagger}_{k y} \hat{b}_{k x} -  \hat{b}^{\dagger}_{k x} \hat{b}_{k y}  \big)\,,
\end{equation}
with $\big[\hat{S},\HM \big] = 0 $.
In the form of \Eq{eq:u1_generator}, however, the symmetry operation $\hat{S}$ involves a hopping  between the two baths in the local bosonic state spaces, which poses a serious impediment for the numerical implementation of the symmetry due to truncation of the bosonic state space. Essentially, the discrete quantum number associated with the symmetry requires a diagonal representation. Hence, it is useful to apply a canonical transformation in order to bring $\hat{S}$ in a diagonal form in the spinor space of $\hat{b^{\dagger}}\equiv(\hat{b}^{\dagger}_x,\hat{b}^{\dagger}_y)$. This leads to
\begin{eqnarray}
\tilde{S} &=& \frac{1}{2} \sigma_z + \sum_k \big( \tilde{b}^{\dagger}_{k y } \tilde{b}_{k y} -  \tilde{b}^{\dagger}_{k x} \tilde{b}_{k x}  \big)\,.
\end{eqnarray}
Note that this transformation also alters the coupling term in the Hamiltonian. In this form, the symmetry sectors are characterized by the $z$-component of the impurity spin and the difference in the bosonic occupation number in both baths in contrast to the hopping term of \Eq{eq:u1_generator}, allowing an exact symmetry implementation in the VMPS procedure in the presence of a truncated bosonic state space.\cite{weichselbaum_QSpace_2012}

Given a simultaneous eigenstate $|q\rangle$ of $\hat{S}$ and $\mathcal{H}$, the application of the generator results in
\begin{equation}
\label{eq:u1quantumnumber}
\tilde{S} |q \rangle = q | q \rangle \quad \text{with } q = \frac{1}{2} \sigma_z + \tilde{N}_y - \tilde{N}_x\,,
\end{equation}
where $\tilde{N}_{i} = \sum_k \tilde{b}^{\dagger}_{ki} \tilde{b}_{ki}$ is the total number of bosons occupying the Wilson chain of the individual baths and $\sigma_z$ is the spin component in $z$-direction. Given any ground state $| G \rangle$, it follows that one may obtain another ground state via $e^{i \phi \tilde{S}}|G\rangle$. Noting that the ground state comes with a symmetric distribution of boson numbers $(\tilde{N}_x=\tilde{N}_y)$, we conclude that $q$ should be chosen to be $\pm1/2$,
\begin{eqnarray}
\tilde{S} |G_{q=\pm1/2}\rangle &=& \pm \frac{1}{2} |G_{q=\pm1/2}\rangle\,, \\
\mathcal{\hat{H}} |G_{q=\pm1/2}\rangle &=& E_g |G_{q=\pm1/2}\rangle\,,
\qph
\end{eqnarray}
where $E_g$ is the ground-state energy. Hence the ground state is doubly degenerated. The expectation value $\langle \sigma_{xy} \rangle$ evaluated using the symmetry ground states  $|G_{\pm1/2}\rangle$ is zero by symmetry. How to reconstruct the magnetization of the symmetry-broken ground state, which is a linear superposition within $|G_{q=\pm1/2}\rangle$, is described in Appendix\,\ref{app2}.

It turns out that the U(1) symmetry implementation cannot be combined with the shifted OBB. Employing a continuous shift $\delta_{k i}$ to the bosonic creation and annihilation operators via \Eqs{eq:b} leads to additional terms of the form $ \delta_{k i}(\tilde{b}_{k i} + \tilde{b}^{\dagger}_{k i})$ in the symmetry generator. These linear corrections add non-diagonal elements to $\tilde{S}$ precluding an explicit implementation of the U(1) symmetry in the way indicated above. This limits the application of symmetry-enforced VMPS effectively to the parameter regime $1/2<s<1$, in which the bosonic state space truncation error does not spoil the calculations of physical quantities such as critical exponents (see Appendix\,\ref{app1} for more details).


\subsection{Energy flow diagrams}
\label{sec:flowdiag}

When VMPS is applied to a Wilson-chain Hamiltonian such as \Eq{eq:hchain},  it is possible to generate an energy-level flow diagram akin to the ones of NRG. To this end, we calculate the eigenvalues $E_k$ of the left block Hamiltonian $\HM^{k}_L$ in each iteration $k<N$ when sweeping from the left to the right end of the Wilson chain truncated to $N$ sites. Multiplied with the proper rescaling factor $\Lambda^{k}$, the spectrum $E^{(k)}_s$ relative to the ground-state energy $E^{(k)}_0=0$ corresponds to the rescaled eigenspectrum determined in a NRG step. The energy flow of excited states is not as smooth as using NRG, since our variational procedure focuses on optimizing the global ground state of the system only. However, it can be systematically improved by incorporating symmetries of the model and keeping more states.

\begin{figure}[t!]
\centering
\includegraphics[width=0.5\textwidth]{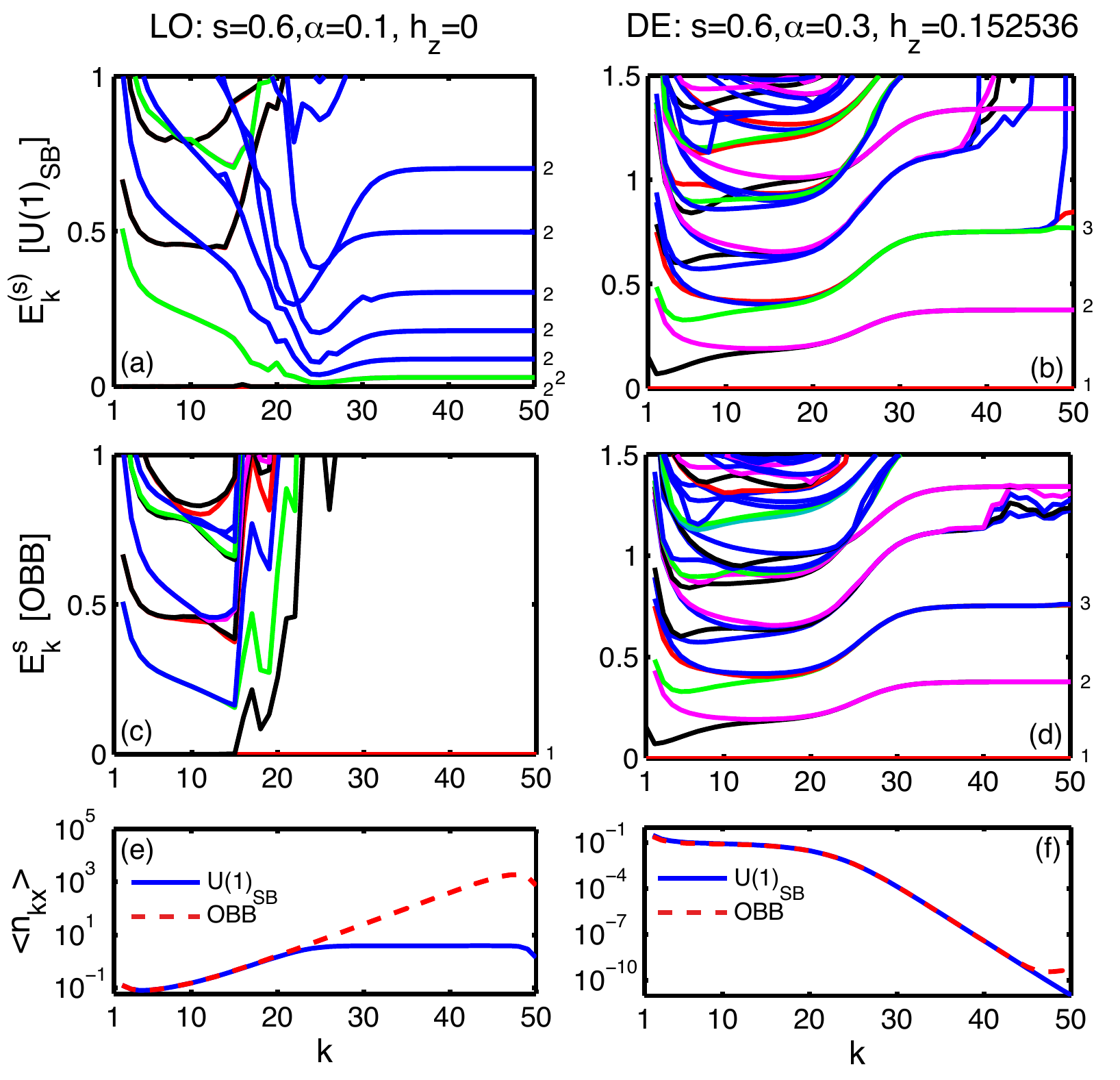}
\vspace{-10pt}
\caption{Characteristic VMPS energy-flow diagrams for {\twobm} with $s=0.6$ in two phases for different values of $\alpha$ and $h_z$. $k$ parameterizes the running energy scale according to $\w = \w_c \Lambda^{-k}$.
While in the two upper panels the flow is generated using the symmetry-enforced VMPS, the center panels show diagrams generated by employing the shifted OBB. The energy levels flow to a localized fixed point in (a),(c) and to a delocalized fixed point in (b),(d) with degenerate (non-degenerate) ground-state space, respectively. The degeneracy of each state is indicated by the numbers to the right side of each curve. The colors in (a) and (b) decode the symmetry label $q$ of each energy level (black and red for $q=\pm1/2$, green and purple for $q=\pm3/2$, and blue for $q\geqslant|5/2|$; matching colors are used in panels (c) and (d)).
Panels (e) and (f) display the corresponding occupation numbers $\langle n_{kx} \rangle$ \eqref{eq:nki}.
}
\label{fig:VMPS_flow}
\end{figure}

Energy flow diagrams contain information about the fixed points of the impurity model, as illustrated in \Fig{fig:VMPS_flow} for {\twobm}, where the upper panels (a,b) are generated by enforcing the U(1) symmetry while for the center panels (c,d) a shifted OBB is employed in the VMPS procedure. The flow towards a localized fixed point with a two-fold generated ground state is depicted in the left panels of \Fig{fig:VMPS_flow}. Only the usage of OBB accounts for the exponential growth of bosonic occupation numbers in the localized phase [cf.~\Fig{fig:VMPS_flow}(e)]. The energy flow in (c) is distorted when introducing the bosonic shift on the Wilson chain, since energy-scale separation is effectively broken due to the exponential growth in local bosonic occupation.  The ground-state degeneracy is conserved, however, when enforcing the symmetry in the VMPS optimization (a). In case the system moves towards a delocalized fixed point with a single ground state at the end of the Wilson chain, both methods generate flow diagrams of similar quality [cf.\ \Figs{fig:VMPS_flow}(b),(d)] since no bosonic shift is necessary to appropriately describe the system's ground state. Hence, energy scale separation remains intact in this case. In the particular example of \Figs{fig:VMPS_flow}(b) and (d), the intermediate fixed point visible at earlier iterations corresponds to the critical fixed point QC2 discussed below.

In addition to determining the system's phase or the convergence of the numerical data, flow diagrams can be used to extract information about the effective energy scales characterizing the crossover between fixed points. For example, the transition from the critical to the DE fixed point is governed by the low-energy scale $T^*\approx \omega_c \Lambda^{-k^{\ast}}$, with $k^{\ast}\approx25$ for the parameters used in \Fig{fig:VMPS_flow}(b) and (d).

\section{Phases and phase diagram}
\label{sec:phases}

In this section, we describe the phase diagram of the U(1)-symmetric {\twobm}, together with the main characteristics of the stable phases.

\subsection{Observables}

The most important observables for {\twobm} employed in this study are the static magnetization,
\begin{equation}
M_\alpha = \frac{1}{2} \langle \sigma_\alpha \rangle \quad (\alpha\equiv x,y,z)\,,
\end{equation}
 and the corresponding susceptibility
\begin{equation}
\chi_\alpha = \lim_{h\to 0} \frac{\partial M_\alpha}{\partial h_\alpha} \quad (\alpha\equiv x,y,z)\,.
\end{equation}
In the case of U(1) symmetry, we distinguish $\chi_{xy}\equiv \chi_{x,y}$ and $\chi_z$ as well as $M_{xy}\equiv M_{x,y}$ and $M_z$.
We will also monitor the occupation numbers of the bath modes of the discretized Wilson chain,
\begin{equation}
\label{eq:nki}
\langle n_{ki} \rangle = \langle \hat{b}_{ki}^{\dagger}\hat{b}_{ki} \rangle
\end{equation}
with $i=x,y$.


\subsection{Stable phases and trivial fixed points}

We start with an overview on the stable phases numerically found for {\twobm}. The description is augmented by an assignment of the corresponding RG fixed points (which are trivial with the exception of the critical phase), with their locations specified in terms of renormalized values of the coupling constants $\alpha$ and $h_z$.

\subsubsection{Free-spin or local-moment phase (F)}

An asymptotically free spin is controlled by the free-spin (F) fixed point, corresponding to vanishing dissipation, $\alpha=0$, and $h_z=0$. The ground state is doubly degenerate, and the susceptibility follows $\chi(T)=1/(4T)$ for all field directions.

\begin{figure}
\centering
\includegraphics[width=.5\textwidth]{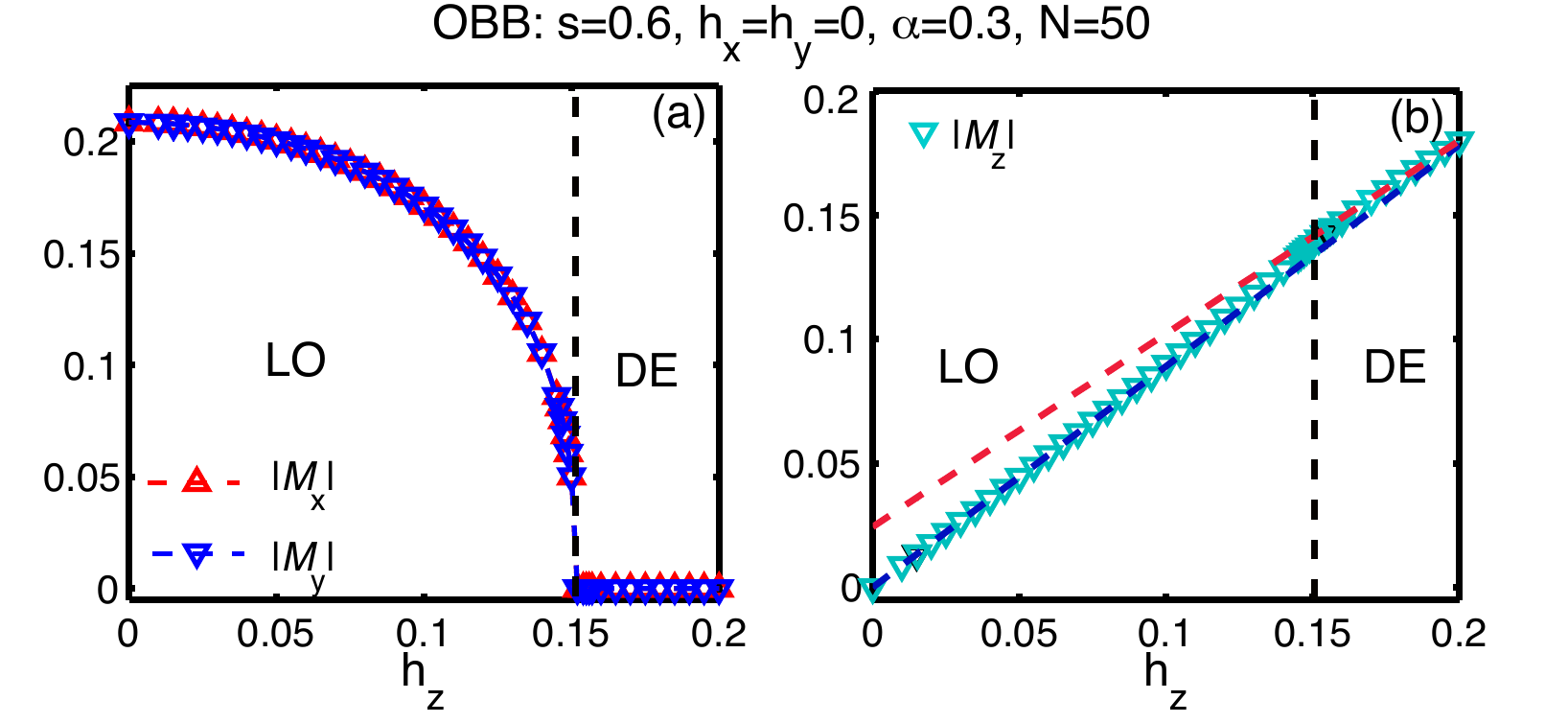}
\caption{Behavior of the magnetization near the LO--DE transition. The order parameter $M_{xy}$ is driven to zero by increasing $h_z$ past the critical value $\hzc$, indicated by the dashed line (a). Correspondingly, the slight kink in the transverse-field response of $M_z$  at the LO--DE transition in (b) indicates the expected higher-order singularity.
We note that the numerics tends to spontaneously favor ordered states with $|M_x|=|M_y|$, as these are the least entangled states.
}
\label{fig:PhaseBound}
\end{figure}

\subsubsection{Localized or strong-coupling phase (LO)}

For large dissipation, the system enters a phase with spontaneously broken U(1) symmetry, controlled by the localized (LO) fixed point. LO is located at $\alpha=\infty$ and $h_z=0$. The bath-oscillator displacements are strongly coupled to the impurity spin, which develops a $T=0$ expectation value in an arbitrary fixed direction in the $xy$-plane, Fig.~\ref{fig:PhaseBound}(a).
This phase is stable for finite (small) transverse field $h_z$ in which case the expectation values of the impurity describe a canted spin, Fig.~\ref{fig:PhaseBound}(b).

Since the symmetry-broken phase exists at $T=0$ only, its associated finite-$T$ susceptibility is expected to be Curie-like, albeit with a classical prefactor,\cite{VBS} $\chi_{xy}(T) = 1/(12T)$.

\subsubsection{Delocalized or polarized phase (DE)}

For dominant transverse field, the impurity spin is polarized along the z-axis and asymptotically decoupled from the bath. This situation is controlled by the delocalized (DE) fixed point, located at $h_z=\infty$ and $\alpha=0$. The ground state is unique, the in-plane magnetizations $M_x$ and $M_y$ vanish, Fig.~\ref{fig:PhaseBound}(a), and all susceptibilities are finite.

\begin{figure}
\centering
\vspace{3pt}
\includegraphics[width=.5\textwidth]{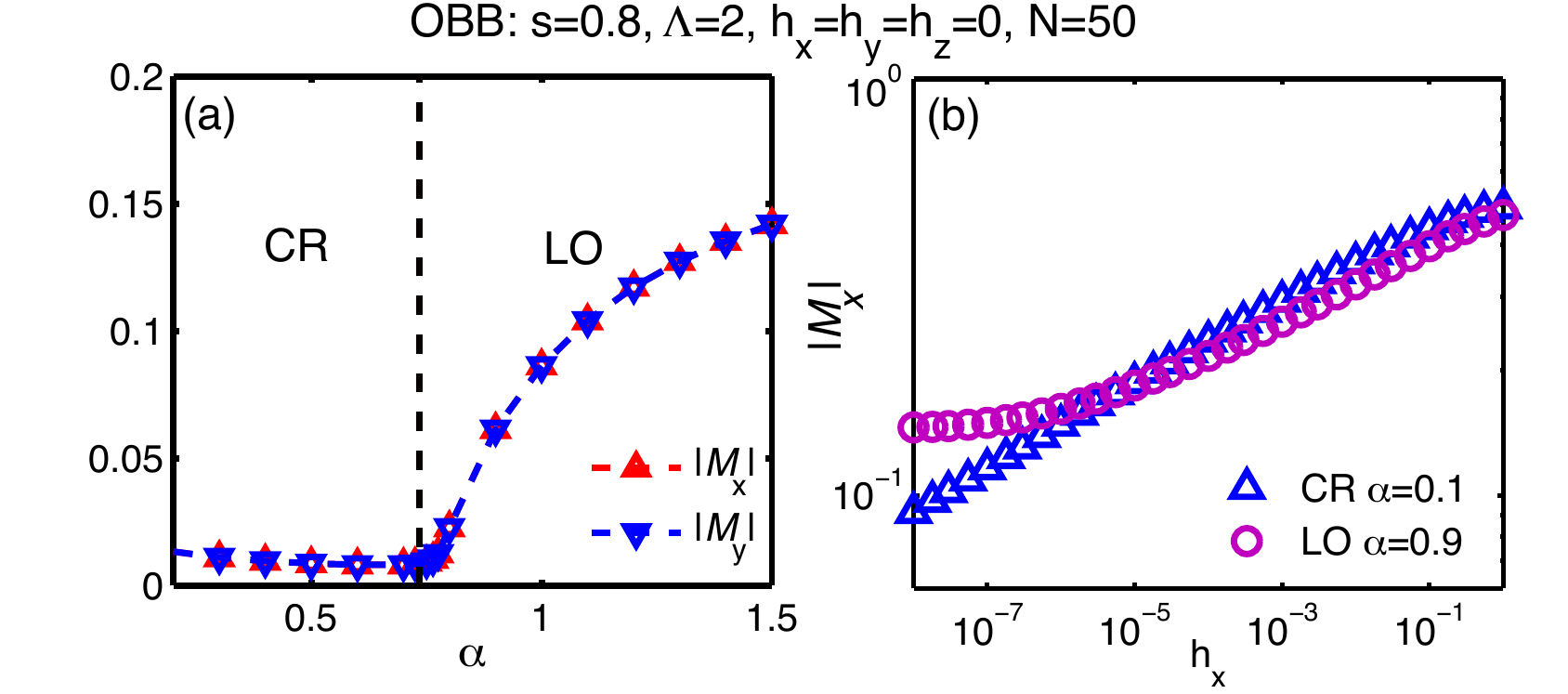}
\caption{Order parameter near the CR--LO transition (a) for different couplings $\alpha$ and (b) response to finite $h_x$ at two points in CR and LO phase. The small but finite magnetization in the CR phase in panel (a) is caused by finite-size effects as discussed in Appendix~\ref{app:finite}.}
\label{fig:CRmag}
\end{figure}

\subsubsection{Critical phase (CR)}

The non-trivial feature of {\twobm} is the existence of a stable critical {\em phase}. This is reached for non-zero (but not too large) dissipation strength $\alpha$ and $h_z=0$ in a certain range of bath exponents $s$. It is controlled by an intermediate-coupling fixed point, not unlike the celebrated two-channel Kondo fixed point.\cite{NB,andrei1984,wiegmann1985}
In this phase, the expectation value of the impurity moment vanishes, but its temporal correlations decay with a fractional power law. This translates into non-linear response functions with fractional exponents, as shown in Fig.~\ref{fig:CRmag}(b).

In contrast to assumptions based on early RG work\cite{anirvan,rg_bfk} -- see also Section~\ref{sec:RGF} below -- the critical phase is {\em not} stable for all dissipation strengths $\alpha$, Fig.~\ref{fig:CRmag}(a), and does not even exist for bath exponents $s<s^\ast$, with a critical value $s^\ast\approx0.76 \pm0.01$.

We note that the critical nature of the CR phase implies significant finite-size effects for the magnetization, as discussed in Appendix~\ref{app:finite}.


\subsection{Numerical determination of phase boundaries}

In order to study the critical phenomena of {\twobm}, it is necessary to accurately determine the phase boundaries, i.e., to numerically calculate the critical coupling $\alpha_c$ and the critical transverse field $\hzc$, which define the location of the LO--CR and LO--DE transitions.

 \begin{figure}
\centering
\includegraphics[width=.5\textwidth]{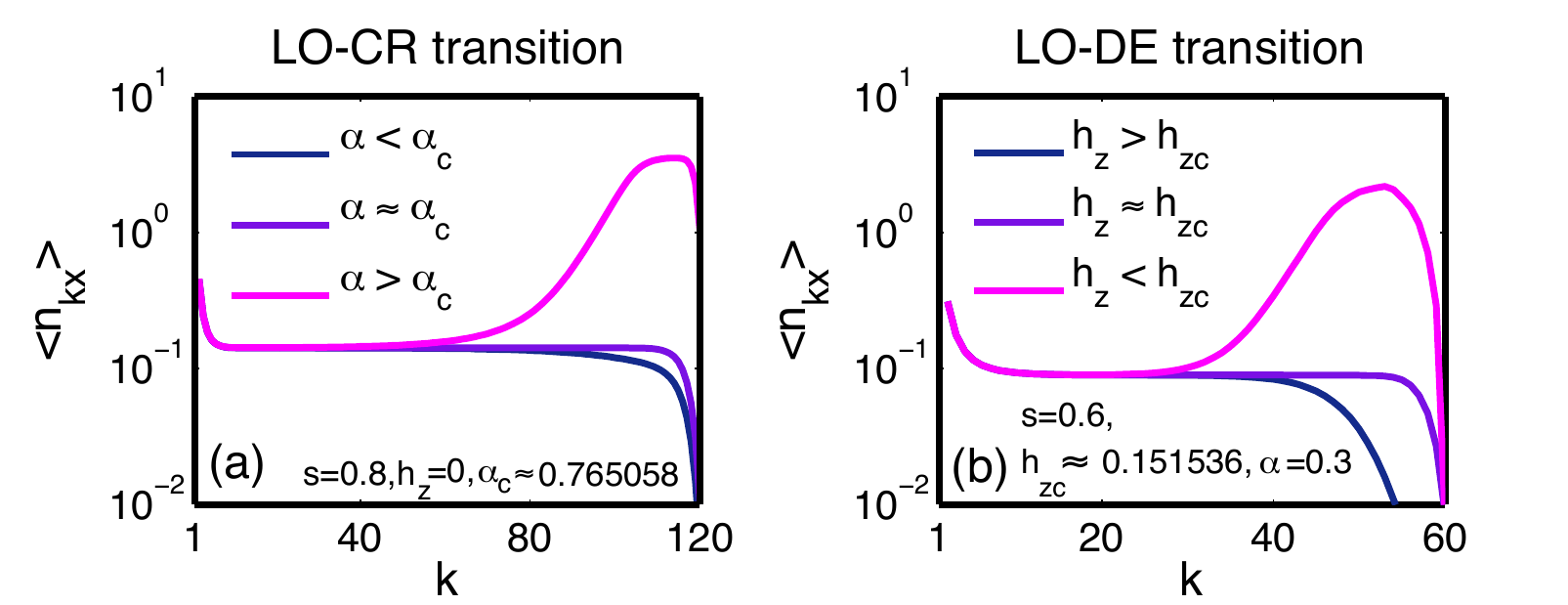}
\caption[Characteristic behavior of the bosonic occupation numbers on the Wilson chain near QC1 (a) and QC2 (b).]{Characteristic behavior of the bosonic occupation numbers on the Wilson chain near QC1 (a) and QC2 (b). In both cases, the occupation numbers stay almost constant throughout the chain directly at the phase boundary, while increasing towards the end of the chain in the localized phase. In the delocalized and critical regime, we observe a steady decay.}
\label{fig:searchPT}
\end{figure}

In our experience, the most accurate and efficient way to calculate $\alpha_c$ and $\hzc$ is to distinguish the phases by the characteristic behavior of the bosonic occupation numbers $\langle n_{ki} \rangle$ on the Wilson chain. The average occupation of boson modes increases towards the end of the Wilson chain in the localized phase, while it decreases in both critical and delocalized phases. Moreover, right at the phase boundary (i.e.\ at criticality) the occupation numbers stay almost constant throughout the chain, except for a sharp decay at the end due to choosing a finite $N$ for the Wilson chain. This characteristic behavior, illustrated in \Fig{fig:searchPT}, can be used to determine the phase boundaries with high accuracy.  We have thus adopted this approach throughout to determine the precise values of $\alpha_c$ and $\hzc$ involved in the results described in section \ref{sec:numerical}. The accessible accuracy depends on the length $N$ of the Wilson chain. Specifically the calculation of $\alpha_c$ or $h_z$ up to $a$ decimals requires a minimal chain length\cite{guo_vmps_2012}
\begin{equation}
\label{eq:accuarcy}
N \propto a\nu \frac{\text{ln}(10)}{\text{ln\,}\Lambda}\,,
\end{equation}
where $\nu$ is the correlation-length exponent. Thus for regions in the phase diagram where $\nu$ becomes larger we have to increase the length of the Wilson chain, making calculations numerically more demanding.


\section{Renormalization-group flow and quantum phase transitions}
\label{sec:flow}

In this section, we use the insights gained in Section~\ref{sec:phases} to deduce the qualitative RG flow of {\twobm}. The discussion will primarily be made in the language and coupling constants of the original Hamiltonian \eqref{eq:h}. A more complete discussion of RG beta functions is given in Section~\ref{sec:RG}.

\begin{figure*}[t!]
\centering
\includegraphics[width=\textwidth]{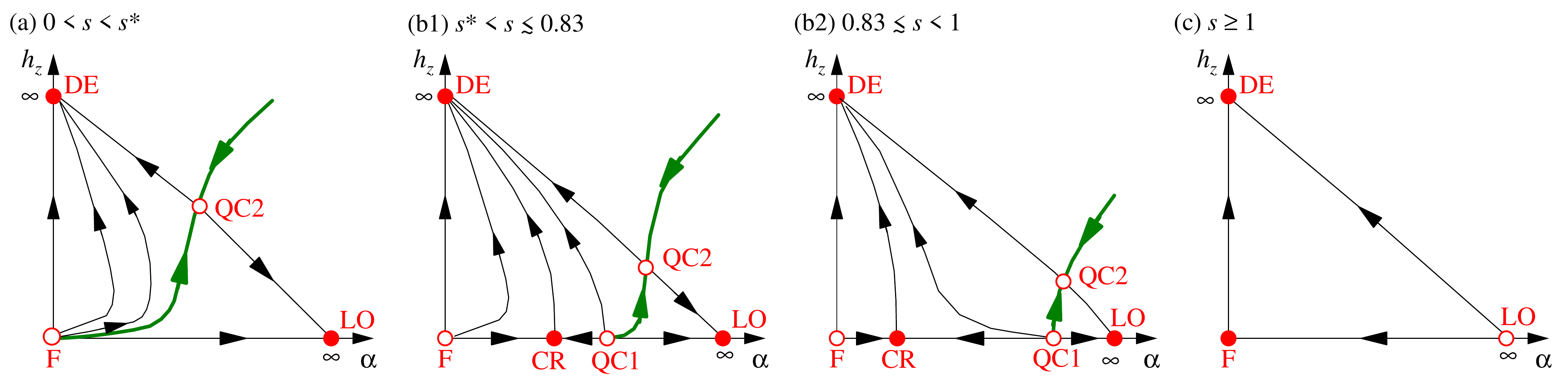}
\caption{
Qualitative RG flow diagrams of the U(1)-symmetric {\twobm} model in a plane spanned by the dissipation strength $\alpha$ and the transverse field $h_z$, as deduced from the VMPS results and supported by the analytical considerations of Section~\ref{sec:RG}. Filled (open) dots denote stable (unstable) RG fixed points; the heavy line is the separatrix corresponding to the DE--LO transition.
Qualitatively distinct behavior is found for the bath-exponent ranges (a) $0<s<s^\ast\approx 0.76$, (b) $s^\ast<s<1$, and (c) $s\geq 1$. The panels (b1) and (b2) illustrate the evolution of both location and relevant-operator dimensions of the fixed points CR and QC1, i.e., $\nu>\nu'$ of QC1 in regime (b1) while $\nu<\nu'$ of QC1 in regime (b2), for details see text.
}
\label{fig:rgflow}
\end{figure*}


\subsection{Qualitative RG flow}

We start by rephrasing our numerical findings in RG language while referring to the qualitative RG flow diagrams in \Fig{fig:rgflow}.

For $h_z=0$ the model {\twobm} displays three phases: F, CR, and LO.
For $s\geq 1$ the free-spin phase F is the only stable phase, i.e., the coupling to the dissipative bath does not qualitatively change the free-spin behavior. This can be contrasted with the physics of {\onebm}, where large dissipation causes localization in the ohmic case $s=1$ -- this distinction reflects the frustration of decoherence in {\twobm}.
For $s<1$, F is unstable against any finite $\alpha$, whereas the localized phase LO is stable for sufficiently large $\alpha$. Finally, the critical phase CR only exists for $s^\ast < s < 1$ and small values of $\alpha$.

A transverse field $h_z\neq0$ destabilizes F for any $s$ and drives the system into the DE phase. CR is unstable against any finite $h_z$ as well. In contrast, LO is stable and hence requires a critical $h_z$ to be destroyed.

This collection allows us to construct the qualitative RG flow diagrams for the ranges of bath exponents $0<s<s^\ast$, $s^\ast<s<1$, and $s\geq 1$, as shown in Fig.~\ref{fig:rgflow}. We also note that the system is always localized for $-1<s\leq0$ provided that $\alpha\neq0$.

In addition to the CR fixed point corresponding to the critical phase, there are two further critical fixed points, QC1 and QC2, which control the quantum phase transitions of {\twobm}. These are described in more detail in the next subsection.


\subsection{Intermediate-coupling fixed points}

For $h_z=0$ there are two fixed points at intermediate coupling, namely CR and QC1, with QC1 controlling the transition between CR and LO. Both intermediate-coupling fixed points are unstable w.r.t. finite $h_z$. Both fixed points only exist for $s^\ast < s < 1$, and it is interesting to discuss their location upon variation of the bath exponent $s$:
As will be shown analytically in Section~\ref{sec:RG} below, CR moves towards F as $s\to 1^-$ whereas QC1 moves towards LO in the same limit, in the fashion characteristic of a lower critical dimension.

In contrast, as $s\to {s^\ast}^+$, both CR and QC1 approach each other, merging at $s=s^\ast$, and disappear for $s<s^\ast$. This merging of two intermediate-coupling fixed points leads to rather unusual behavior, with the phase boundary of LO jumping upon variation of $s$ across $s^\ast$.

For finite $h_z$ a transition can be driven between DE and LO, which is controlled by QC2. QC2 moves towards LO as $s\to1^-$, again in a manner of a lower critical dimension. This is consistent with the fact that the localized phase ceases to exist for $s>1$. In the limit $s\to0^+$, QC2 approaches DE, such that DE becomes unstable w.r.t. finite $\alpha$ for $s\leq0$, reflecting that the system is always localized.


\subsection{Critical exponents}
\label{sec:exponents}

The quantum phase transitions of {\twobm} can be characterized by standard critical exponents.\cite{goldenfeld} For a transition which can be accessed by varying $\alpha$ (at fixed $h_z$), with the transition point at $\alpha=\alpha_c$, the following exponents can be defined from the zero-temperature order parameter $M_{xy}$ and its conjugate field $h_{xy}$:
\begin{align}
M_{xy}(\alpha, h_{xy}=0) &\propto (\alpha-\alpha_c)^\beta\,,\label{eq:beta}\\
M_{xy}(\alpha=\alpha_c, h_{xy}) &\propto h_{xy}^{1/\delta}\,. \label{eq:delta}
\end{align}

Transitions which occur at finite $h_z$ can also be driven by varying $h_z$ at fixed $\alpha$;
correspondingly, the exponent $\beta$ may be defined via $M_{xy} \propto (\hzc-h_z)^\beta$ as well.
In contrast, for $h_z=0$ transitions, $h_z$ takes a role different from $(\alpha-\alpha_c)$, as it
reduces the symmetry of the model from U(1)\,$\times$\,Z$_2$ to U(1). It is useful to introduce an
exponent for the non-linear response to $h_z$ according to
\begin{equation}
\label{deltapr}
M_{z}(\alpha=\alpha_c, h_{xy}=0, h_{z}) \propto h_{z}^{1/\delta'}\,.\\
\end{equation}

A correlation-length exponent is defined as usual from the divergence of a correlation length, here
equivalent to the vanishing of a crossover energy $T^\ast$ according to
\begin{equation}
T^\ast(\alpha, h_{xy}=0) \propto |\alpha-\alpha_c|^\nu\,; \label{nu}\\
\end{equation}
note that there is no separate dynamical exponent for the $(0+1)$-dimensional impurity model under consideration, formally $z=1$.
For fixed points located at $h_z=0$ which are unstable towards finite $h_z$ we additionally define
\begin{equation}
\label{nupr}
T^\ast(\alpha=\alpha_c, h_{xy}=0, h_z) \propto |h_z|^{\nu'}\,.\\
\end{equation}

The linear-response order-parameter susceptibility diverges at the quantum critical point, in the approach from either smaller $\alpha$ or from finite $T$, according to
\begin{align}
\chi_{xy}(\alpha, T=0) &\propto (\alpha_c-\alpha)^{-\gamma} \,\\
\chi_{xy}(\alpha=\alpha_c, T) &\propto T^{-x} \,.
\label{xdef}
\end{align}
Within the quantum-to-classical correspondence, $x$ is related to the finite-size scaling of the
classical model's susceptibility at criticality. One may also consider the dynamic version of the order-parameter susceptibility, which follows a power-law behavior at criticality,
\begin{equation}
\chi_{xy}(\alpha=\alpha_c, \w) \propto \w^{-y} \,,
\end{equation}
corresponding to power-law autocorrelations of the impurity spin in time. The exponent $y$ contains the same information as the usually-defined anomalous exponent $\eta$, with $y\equiv 2-\eta$. At the critical points of {\twobm} (and other spin models with long-ranged interactions), $\eta=2-s$ (equivalently, $y=s$) is believed to be an exact relation, see also Section~\ref{sec:RG}.

Due to the anisotropic nature of the spin fluctuations, different power laws arise for the $z$-component susceptibility:
\begin{align}
\chi_{z}(\alpha=\alpha_c, T) &\propto T^{-x'} \,,\\
\chi_{z}(\alpha=\alpha_c, \w) &\propto \w^{-y'} \,.
\end{align}

Finally, it is also useful to introduce exponents which describe the location of the DE--LO phase boundary at small $h_z$. For $0<s<s^\ast$ this phase boundary is connected to the $\alpha=h_z=0$ point, and we define
\begin{equation}
\label{psidef1}
\hzc \propto \alpha^\psi\,.
\end{equation}
In contrast, for $s^\ast<s<1$ the DE--LO boundary terminates at the CR--LO transition located at $\alpha=\alpha_c$ and $h_z=0$, and we use
\begin{equation}
\label{psidef2}
\hzc \propto (\alpha-\alpha_c)^\psi\,.
\end{equation}



\subsection{Scaling}
\label{sec:scaling}
The exponents introduced above can be related to each other via scaling relations, following textbook strategy.\cite{goldenfeld} The standard scaling relations do hold,
\begin{align}
\beta\,\delta &= \beta + \gamma\,, \label{s1} \\
\gamma &= (2-\eta) \nu \equiv y \nu\,. \label{s2}
\end{align}
The exact result $y=s$ then implies
\begin{equation}
\gamma = s\nu\,. \label{s3}
\end{equation}
For critical points with hyperscaling, additional scaling relations apply, which involve spatial dimensionality $d$:
\begin{align}
2\beta + \gamma &= \nu d\,, \label{hs1}\\
\delta &= \frac{d+2-\eta}{d-2+\eta}\, .\label{hs2}
\end{align}
Furthermore, hyperscaling implies $x=y$. For $d=1$ and using the exact result $y=s$, the hyperscaling relations can be converted into
\begin{align}
x &= s\,,\label{hs3}\\
\beta &= \gamma \frac{1-s}{2s} = \nu \frac{1-s}{2}\,, \label{hs4}\\
\delta &= \frac{1+s}{1-s}\, .\label{hs5}
\end{align}

For critical points of {\twobm} with $h_z=0$, the scaling hypothesis underlying hyperscaling can be extended to include the dependence on $h_z$ [in addition to that on $(\alpha-\alpha_c)$, $h_{xy}$, and $T$]. This then yields additional hyperscaling relations: $x'=y'$ and
\begin{align}
\nu' &= 1 + \frac{1}{\delta'}\,, \label{eq:hyp_nup} \\
\delta' &= \frac{1+x'}{1-x'}\,,\label{hs6}
\end{align}
see Appendix~\ref{app:hyper} for a derivation.

We recall that hyperscaling, which is of general interest because it implies simple and powerful scaling relations which can be applied in analyzing both experimental and numerical data, usually holds for phase transitions below their upper critical dimension. Hyperscaling is spoiled by the existence of dangerously irrelevant variables in the critical theory; the most important example here is the quartic coupling of a (classical) $\phi^4$ theory in dimensions $d>4$.


\section{Epsilon expansions and critical behavior}
\label{sec:RG}

We now describe analytical approaches to the critical-point properties of {\twobm}, utilizing the
field-theoretic toolbox with renormalization-group and epsilon-expansion techniques.


\subsection{Expansion around F: CR phase}
\label{sec:RGF}

The free-impurity fixed point F is characterized by a doubly-degenerate impurity at $\alpha=0$, $h_z=0$. Tree-level power counting yields the scaling dimensions (recall that $\alpha\propto\lambda_{qi}^2$):
\begin{align}
{\rm dim}[\alpha] &= 1-s\,, \\
{\rm dim}[h_z] &= 1\,.
\end{align}

\subsubsection{DE--LO phase boundary}

From the scaling dimensions one can immediately read off the asymptotic behavior of the flow trajectories leaving the F fixed point, $h_z \propto \alpha^{1/(1-s)}$.
This also applies to the DE--LO separatrix in the exponent range $0<s<s^\ast$, yielding the phase-boundary exponent according to Eq.~\eqref{psidef1} as
\begin{equation}
\label{psires1}
\psi = \frac{1}{1-s}\,.
\end{equation}

\subsubsection{RG analysis}

Now we turn to a RG analysis of the flow of $\alpha$ at $h_z=0$. Given that the dissipation is a marginal perturbation at $s=1$, this is akin to a standard epsilon expansion with $\epsilon = 1-s$, which can give reliable results for small $(1-s)$.
Straightforward perturbation theory, along the lines of Refs.~\onlinecite{VBS,rg_bfk,kv04}, yields the two-loop beta function:\cite{rg_bfk,ren_foot}
\begin{align}
\label{betaF}
\beta(\alpha) &= (1-s) \alpha - \alpha^2 + \alpha^3\,.
\end{align}
This beta function indicates the existence of an infrared-stable fixed point at
\begin{equation}
\label{alstar}
\alpha^\ast = (1-s) + (1-s)^2 + \mathcal{O}((1-s)^3)
\end{equation}
and $h_z=0$ -- this is the CR fixed point.
Its properties can be obtained in a double expansion in $\alpha$ and $(1-s)$.
The exact result $x=y=s$ follows from the diagrammatic structure of the susceptibility\cite{VBS} or, alternatively, from a Ward identity.\cite{rg_bfk}
From this we have:
\begin{align}
1/\delta  &= \frac{1-s}{1+s}\label{eq:delta_cr}
\end{align}
as above. The $z$ component correlator requires an explicit computation, with the two-loop result:\cite{rg_bfk}
\begin{equation}
1-y' = 2(1-s) + (1-s)^2 + \mathcal{O}((1-s)^3) \,.
\end{equation}
The remaining exponents involving the $h_z$ response can be calculated from the hyperscaling relations \eqref{eq:hyp_nup} and \eqref{hs6}, with the result:
\begin{align}
\label{eq:nup_cr}
1/\nu'    &= s - \frac{(1-s)^2}{2} + \mathcal{O}((1-s)^3) \,, \\
1/\delta' &= 1-s + \frac{3(1-s)^2}{2} + \mathcal{O}((1-s)^3)\,.\label{eq:deltap_cr}
\end{align}

We point out that the RG flow towards the CR fixed point is rather slow, because the leading irrelevant operator, its prefactor being $(\alpha-\alpha^\ast)$, has a small scaling dimension of $\w=1-s$. Therefore, quickly converging numerical results are best obtained using a bare coupling close to $\alpha^\ast$.\cite{ren_foot}

\subsubsection{Disappearance of CR for $s<s^\ast$}

It is interesting to note that the beta function in Eq.~\eqref{betaF} displays {\em two} non-trivial fixed points at $\alpha^\ast_{1,2} = 1/2 \pm \sqrt{1/4 - (1-s)}$. While $\alpha^\ast_2$ corresponds to the stable CR fixed point of Eq.~\eqref{alstar}, the infrared-unstable fixed point at $\alpha^\ast_1$ is outside the range of validity of the epsilon expansion. However, {\em if} we choose to ignore this restriction, the comparison with the numerical results suggests to associate $\alpha^\ast_1$ with QC1. Remarkably,
$\alpha^\ast_1$ and $\alpha^\ast_2$ approach each other upon decreasing $s$ from unity, and the criterion $\alpha^\ast_1=\alpha^\ast_2$ yields $s^\ast=3/4$ which is extremely close to the numerical determined value of $s^\ast\approx0.76\pm0.01$ where CR and QC1 merge.

While this can be interpreted as a remarkable success of the epsilon expansion -- it predicts not only the existence of the CR phase, but also its disappearance for $s<s^\ast$ -- we note that this epsilon expansion does not provide means to reliably calculate critical properties of QC1, simply because $\alpha^\ast_1$ is never small. As we show below, the presence (absence) of hyperscaling in a field at CR (QC1) even indicates a qualitative difference between CR and QC1 which is not apparent from this epsilon expansion.


\subsection{Expansion around DE: QC2}
\label{sec:RGD}

It is also possible to devise an expansion around the delocalized fixed point DE, located at $h_z=\infty$, $\alpha=0$. Such an expansion has been first used in Ref.~\onlinecite{VTB} for {\onebm}, but the analysis there missed the presence of a dangerously irrelevant operator (the quartic coupling $u$ in Eq.~\eqref{heff} below) and erroneously assumed hyperscaling, which led to partially incorrect conclusions.\cite{VTB_err} Here we correct this approach and apply it to {\twobm}.
For convenience we assume equal couplings between the impurity and the different oscillator modes, $\lambda_{qi} \equiv \lambda_i$, such that the energy dependence of $J_i(\omega)$ is contained in the density of states of the oscillator modes $\omega_q$, and we have $\alpha_i \propto \lambda_i^2$.

\subsubsection{Projection}

\begin{figure}[t]
\centering
\includegraphics[width=.48\textwidth]{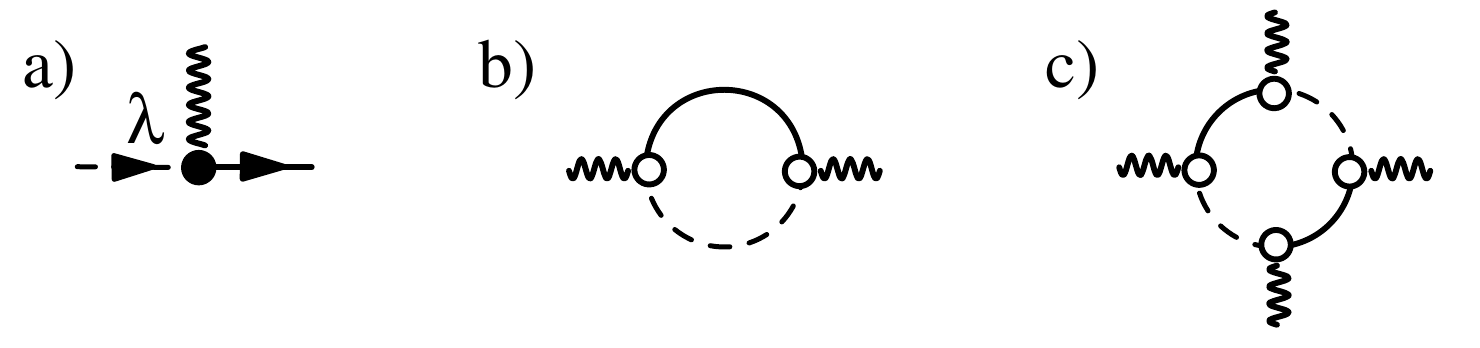}
\caption[]{Feynman diagrams occurring in the perturbation expansion around DE.
Full/dashed lines denote the propagators of the $|\uparrow\rangle$ and $|\!\downarrow\rangle$
impurity states -- the two states are separated by a gap $h_z$.
The wiggly line is the local bath boson $\phi_{x,y}$.
a) Interaction vertex $\lambda$.
b) Bilinear $\phi$ term.
c) Quartic $\phi$ term.
}
\label{fig:diag1}
\end{figure}

At DE we have a single low-lying impurity level, $|\uparrow\rangle$, while $|\downarrow\rangle$ is separated by an energy $h_z$. Low-energy interaction processes between the impurity and the baths arise in second-order perturbation theory, controlled by the coupling
\begin{equation}
\kappa_i = \lambda_i^2/h_z \,.
\end{equation}
In the low-energy sector -- this corresponds to projecting out the $|\downarrow\rangle$ state -- the effective theory reads (assuming from here on $\alpha_x=\alpha_y$ or $\kappa_x=\kappa_y$):
\begin{equation}
\label{heff}
\HM_\eff = \HM_{\bath} + m(\phi_x^2 + \phi_y^2) + u (\phi_x^2+\phi_y^2)^2
\end{equation}
with $m=-\kappa$ and $u=\kappa^2$. We have defined
\begin{equation}
\phi_i = \sum_q ( \hat{a}^\pdag_{q i} + \hat{a}_{q i}^{\dagger} ) \,,
\end{equation}
and we have omitted higher-order terms in Eq.~\eqref{heff}. Fig.~\ref{fig:diag1} illustrates how the $m$ and $u$ terms are generated from $\mathcal{H}$ of the original model {\twobm}; this approach is valid provided that $\lambda\ll h_z,\w_c$.

\subsubsection{Local $\phi^4$ theory and quantum-to-classical correspondence}

The theory $\HM_\eff$ can be understood as a theory for the local bosonic fields $\phi_{x,y}$. Their ``bare'' propagator arises from $\HM_{\bath}$ and is given by $G_\phi^{-1}(i\nu_n) = i A_0 {\rm sgn}(\nu_n) |\nu_n|^s + A_1$ at low energies, with $A_1 = -2\w_c/s$ for the power-law spectrum in Eq.~\eqref{power}. The main role of the impurity in $\HM_\eff$ is that of an additional mass term (recall that the impurity spin degree of freedom has been projected out).

Importantly, $\HM_\eff$ in Eq.~\eqref{heff} is identical to a local $\phi^4$ theory for an XY-symmetric order-parameter field $(\phi_x,\phi_y)$ with long-ranged interactions $\propto 1/\tau^{1+s}$ in imaginary time.
It displays a critical point which corresponds to a vanishing $\phi$ mass. Ignoring the influence of the quartic interaction $u$, this happens at $m_c=A_1 < 0$; this can alternatively be understood by interpreting $m$ as the strength of a potential scatterer, where $m_c G_\phi(0) = m_c/A_1 = 1$ is the condition for a zero-energy pole of the T matrix.
For positive mass, i.e., small $\kappa$, $\HM_\eff$ is in a disordered phase corresponding to DE, whereas negative mass drives the system into an ordered phase with spontaneously broken XY symmetry -- this can be identified with LO.

Consequently, the critical point of $\HM_\eff$ corresponds to QC2. As the $\phi^4$ theory in question is the low-energy theory of a classical XY chain with long-range interactions, we conclude that QC2 obeys a quantum-to-classical correspondence at least if QC2 is located in the small-$\kappa$ parameter regime where the above mapping to $\HM_\eff$ is valid, i.e., for small $s$. The critical properties for the classical XY chain are listed in Section~\ref{sec:xyexp} below; in particular, mean-field behavior obtains for $s<1/2$.

\subsubsection{RG analysis}

An alternative approach to $\HM_\eff$ is to analyze the flow of the couplings $m$ and $u$ near the DE fixed point by RG means. Power counting w.r.t. the $\lambda=0$ limit gives
\begin{align}
{\rm dim}[m] &= -s\,,\\
{\rm dim}[u] &= 2s-1\,,
\end{align}
i.e., $m$ is marginal at $s=0$ while $u$ is irrelevant.
Near $s=0$ we can follow the flow of $m$ which yields at one-loop order:
\begin{eqnarray}
\label{mast}
\beta(m) = -s m + m^2 \,.
\end{eqnarray}
Besides the stable DE fixed point at $m=0$ (i.e.\ $\alpha=0$) this
flow equation displays an infrared unstable fixed point (QC2) at
\begin{equation}
m^\ast = s + \mathcal{O}(s^2)
\label{fp}
\end{equation}
which controls the transition between the DE and LO phases. Corrections from $u$ only enter at higher orders in $s$, because the initial values obey $u=m^2$. We note that the value of $m^\ast$ in Eq.~\eqref{mast} is consistent with $m_c=-A_1$ from above.
Expanding the RG beta function around the fixed point (\ref{fp}) gives the correlation-length exponent
\begin{equation}
1 / \nu = s + \mathcal{O}(s^2) \,,
\label{nude}
\end{equation}
apparently in agreement with the classical mean-field result \eqref{eq:numf}.

One may employ renormalized perturbation theory to calculate critical exponents in a double expansion in $m$ and $s$. This is, however, complicated by the facts that (i) for many observables of the original model {\twobm} one needs to restore the impurity Hilbert space, i.e., undo the elimination of the $|\downarrow\rangle$ state, and (ii) the quartic coupling $u$ is dangerously irrelevant and cannot be neglected. For selected exponents, we have checked that this procedure yields results consistent with Eqs.~\eqref{eq:etamf}-\eqref{eq:xmf}.


\subsection{Quantum-to-classical mapping of {\twobm}}
\label{sec:QCC}

One may ask whether a general mapping of {\twobm} to a classical statistical-mechanics model exists. Such a mapping, using a Feynman path integral representation, can indeed be formulated for the single-bath spin-boson model ({\onebm}) and directly leads to an Ising chain with both short-ranged and long-ranged $1/r^{1+s}$ interactions.\cite{leggett,vojta_NRG_2012}

Here we sketch what happens when applying the same procedure to {\twobm}. For simplicity, we restrict ourselves to $\vec{h}=0$. The Hamiltonian may be written as $\HM = \HM_{x} + \HM_{y} + \HM_{\bath}$ with
\begin{align}
\HM_{i} &= \sum_q  \lambda_{q i} \frac{\sigma_i}{2} Q_{qi} ~~(i=x,y), \notag\\
\HM_{\bath} &= \sum_{i = x,y} \sum_{q} \left(\frac{P_{qi}^2}{2 m_{q}} + \frac{m_{q} \w_q^2 Q_{qi}^2}{2}\right).
\end{align}
The Feyman path integral for the partition function can be expressed using eigenstates of $\sigma_x$, $\sigma_y$, and the oscillators coordinates. Inserting the identities for the spin variables -- those for the oscillator coordinates are standard and do not lead to any complications -- it reads:
\begin{align}
Z &= {\rm Tr}_{P,Q} \int \mathcal{D}\sigma_x \mathcal{D}\sigma_y
\langle {\sigma_x}_N | e^{-\epsilon \HM_x} | {\sigma_y}_{N-1}\rangle \notag\\
&\langle {\sigma_y}_{N-1} | e^{-\epsilon \HM_y} | {\sigma_x}_{N-1} \rangle e^{-\epsilon \HM_{\bath}} \langle {\sigma_x}_{N-1} | \ldots | {\sigma_x}_0 \rangle
\label{fpi}
\end{align}
where $N$ is the number of Trotter slices, $\epsilon=\beta/N$, and ${\sigma_x}_0 = {\sigma_x}_N$.
In principle, a classical spin model can now be obtained by integrating out the bath oscillators, which generates long-ranged interactions for the variables coupled to these oscillators. In the case of {\twobm}, these are both $\sigma_x$ and $\sigma_y$, such that one ends up with a representation in terms of sets of Ising spins. To rewrite this in terms of a classical spin model requires to express the matrix elements in Eq.~\eqref{fpi} as exponentials of classical interactions. Remarkably, the set of matrix elements $\langle \sigma_x|\sigma_y\rangle$ cannot be expressed as $e^{H_c(\sigma_x,\sigma_y)}$ with a classical real Hamiltonian function $H_c$, i.e., the Feynman path-integral representation of {\twobm} leads to an ill-defined classical model with negative Boltzmann weights.\cite{tv_pc} Clearly, this problem can be traced back to the non-commutativity of the two spin components which couple to the oscillator baths in {\twobm}.

We recall, however, that the physics of the QC2 fixed point of {\twobm} {\em can} be mapped onto that of a classical XY model at least near $s=0$, see Section~\ref{sec:RGD} above. Assuming that the character of QC2 does not change fundamentally as function of $s$, this implies that a quantum-to-classical correspondence indeed holds for QC2. As will be shown in Section~\ref{sec:numerical} below, our numerical results for the critical behavior near QC2 are perfectly consistent with this assertion.


\subsection{Exponents of classical XY chain}
\label{sec:xyexp}

Here we collect and summarize the available results for critical exponents of the classical XY chain
with long-range interactions decaying as $1/r^{1+s}$ -- these have been discussed in
Refs.~\onlinecite{fisher_critical_1972,koster}. The classical model has a thermal phase transition for
$0<s<1$; no ordered phase exists for $s\geq 1$.

For $s<1/2$ and $s\gtrsim 1/2$ one may utilize the language of a $\phi^4$ theory. Power counting shows
that the quartic interaction is marginal for $s=1/2$, such that mean-field behavior attains for
$s<1/2$, with
\begin{align}
1/\nu &= s\,, \label{eq:numf}\\
\eta &= 2-s\,, \label{eq:etamf}\\
\beta & = 1/2\,, \label{eq:betamf}\\
\gamma &= 1\,,\\
\delta &= 3\,,\\
x &= 1/2\,, \label{eq:xmf}
\end{align}
with hyperscaling being violated.

In the non-mean-field regime, $s>1/2$, one can obtain exponents in an expansion in $\epsilon = s-1/2$,
with two-loop results as quoted in Ref.~\onlinecite{fisher_critical_1972}:
\begin{align}
\gamma &= 1 + \frac{8}{5} \epsilon - \frac{16}{25}\left(1-\frac{17\mathcal{A}(1/2)}{5}\right) \epsilon^2 + \mathcal{O}(\epsilon^3) \,, \\
\eta &= 2-s\,,
\end{align}
the latter result is believed to be exact to all orders,\cite{fisher_critical_1972,suzuki} and the constant
\begin{equation}
\mathcal{A}(s) = s [\psi(1) - 2\psi(s/2) + \psi(s)]
\end{equation}
in terms of the digamma function $\psi(x)$, with $\mathcal{A}(1/2) = 2.957$. Hyperscaling holds for
$1/2 < s < 1$; this allows to derive the remaining exponents:
The correlation-length exponent $\nu$ follows from the scaling relation $\gamma = (2-\eta)\nu$ \eqref{s2}, while
$\beta$ can be read off from the hyperscaling relation $\beta = \gamma (1-s)/(2s)$ \eqref{hs4}, with the results
\begin{align}
1/\nu   &= 1/2 + 1/5 \epsilon - 3.217 \epsilon^2 + \mathcal{O}(\epsilon^3) \,,\label{eq:nu_rg}\\
1/\beta &= 2 + 24/5 \epsilon - 3.269  \epsilon^2 + \mathcal{O}(\epsilon^3) \,. \label{eq:beta_rg}
\end{align}

Near $s=1$ RG equations can be derived\cite{koster} using a variant of a method proposed by Polyakov
\cite{polyakov75} -- this is similar to an ordered-phase expansion in $(2+\epsilon)$ dimensions for
magnets with short-range interactions. Exponents are formally obtained in an expansion in $(1-s)$; the
one-loop results read:\cite{koster}
\begin{align}
1/\nu &= 1-s + \mathcal{O}((1-s)^2) \label{eq:nu_rg2}\,,\\
\eta &= 2-s\,,
\end{align}
the latter result again believed to be exact to all orders. Using hyperscaling, we obtain the one-loop
result for $\beta$ as
\begin{equation}
\beta = 1/2 + \mathcal{O}((1-s)^2)\,. \label{eq:beta_hyp}
\end{equation}


 \begin{figure*}
\centering
\includegraphics[width=1\textwidth]{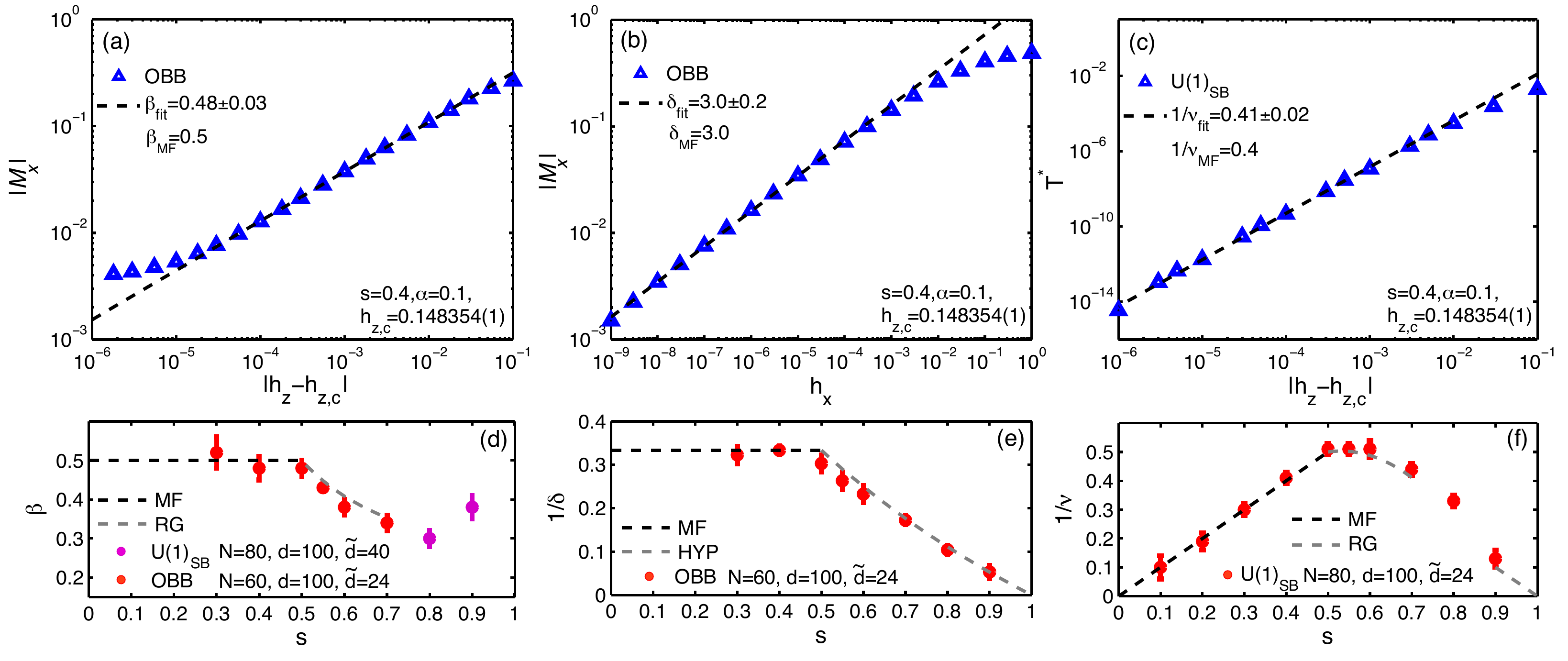}
\vspace{-15pt}
\caption{VMPS results for critical exponents $\beta$, $\delta$ and $\nu$ at the LO--DE quantum phase transition. Analogous to \Fig{fig:exp_qc1}, the upper panels (a-c) display the calculated VMPS results for the order parameter and the crossover scale close to/at the critical point for $s=0.4$, respectively. The $s$-dependent behavior of the critical exponents $\beta, \delta$ and $\nu$ obtained from the respective power-law fits is illustrated in lower panels (d-f). In addition, these panels contain the corresponding predictions of the classical XY model (dashed lines), which we find to be in excellent agreement with the numerical data.}
\label{fig:exp_qc2}
\end{figure*}

\section{Numerical results for critical exponents}
\label{sec:numerical}

Taking into account the insights gained in the preceding Sections~\ref{sec:flow} and \ref{sec:flow}, we now focus on the numerical results obtained for the critical behavior of {\twobm}.
To this end, we employ the VMPS methodology as introduced in Section~\ref{sec:method} at the quantum phase transitions QC1 and QC2, as well as in the CR phase to extract various critical exponents.

Our main results are that (i) the transition between LO and DE, controlled by QC2, indeed obeys quantum-to-classical correspondence, i.e., its critical properties are that of a classical XY chain with long-range interactions, and (ii) the transition between CR and LO, controlled by QC1, is of unusual nature, with no quantum-to-classical correspondence and hyperscaling present only at $h_z=0$.

In the following, we distinguish VMPS results obtained using the shifted OBB  (denoted by OBB) from the symmetry-enforced approach (denoted by U(1)$_{\mathrm{SB}}$). In all calculations, we work with a fixed Wilson discretization parameter $\Lambda=2$, bond dimension $D=60$ and local bosonic dimension $d_k=100$ while varying the chain length $N$ and the effective local dimension $\tilde{d}_k \leqslant d_k/2$, as denoted in the Figures. Since $d_k$ and $\tilde{d}_k$ are set equal for different sites during a single VMPS calculation, we omit the label $k$ in the following. This choice of $D$ and $\tilde{d}$ ensures that we keep all singular values larger than $10^{-5}$ in our calculation.


\subsection{Transition between LO and DE phase}

We start the discussion with the continuous quantum phase transition between the LO and DE phases that is controlled by the critical fixed point QC2. As explained in Section~\ref{sec:RG}, this transition should correspond to the thermal transition of the XY chain with long-ranged interactions. Here we show that our numerical results are in excellent agreement with analytical predictions of scaling and epsilon-expansion calculations, listed in Section~\ref{sec:xyexp}, and therefore fully confirm the quantum-to-classical correspondence.

\subsubsection{Order parameter exponent $\beta$}

Accessible only at finite $h_z$, we drive the transition between the DE and LO phases by varying $h_z$ for fixed $\alpha$. Hence, the critical exponent $\beta$ is defined via $M_{xy} \propto (\hzc - h_z)^{\beta}$ at the phase boundary moving into the LO phase. Panels (a) and (d) of \Fig{fig:exp_qc2} show the corresponding numerical data. The characteristic power-law behavior of the magnetization for fixed $s=0.4$ close to the critical point on a log-log scale is displayed in \Fig{fig:exp_qc2}(a). The exponent derived from a linear fit to this data, namely  $\beta=0.48\pm0.03$, corresponds to the mean-field prediction in \Eq{eq:betamf} within the error bars. Deviations from power-law behavior at small $|h_z-\hzc|$ can be attributed to a combination of finite chain length $N\leqslant 80$ and numerical errors of VMPS. Our numerical method generates power-law plots of similar quality for all $s\geqslant0.3$; the resulting exponents are collected in \Fig{fig:exp_qc2}(d). These are found to be in excellent agreement with the predictions of the quantum-to-classical mapping. As for the classical XY chain, the exponent assumes its mean-field value  $\beta=1/2$ for $0<s<1/2$, while it follows the two-loop RG results in \Eq{eq:beta_rg} for $s=1/2+\epsilon$. In the limit of $s\to1^-$, $\beta$ shows the tendency to approach the value 1/2, consistent with \Eq{eq:beta_hyp}. The growing shifts in the localized phase, in combination with the decreasingly low-energy scale necessary to precisely access the critical point, prevents our numerics to extract accurate results for $\beta$ in the deep subohmic regime ($s\leqslant 0.3$). The second issue (decreasingly low-energy scale) also applies in the limit $s\to1^-$.

\begin{figure}[b]
\centering
\includegraphics[width=0.5\textwidth]{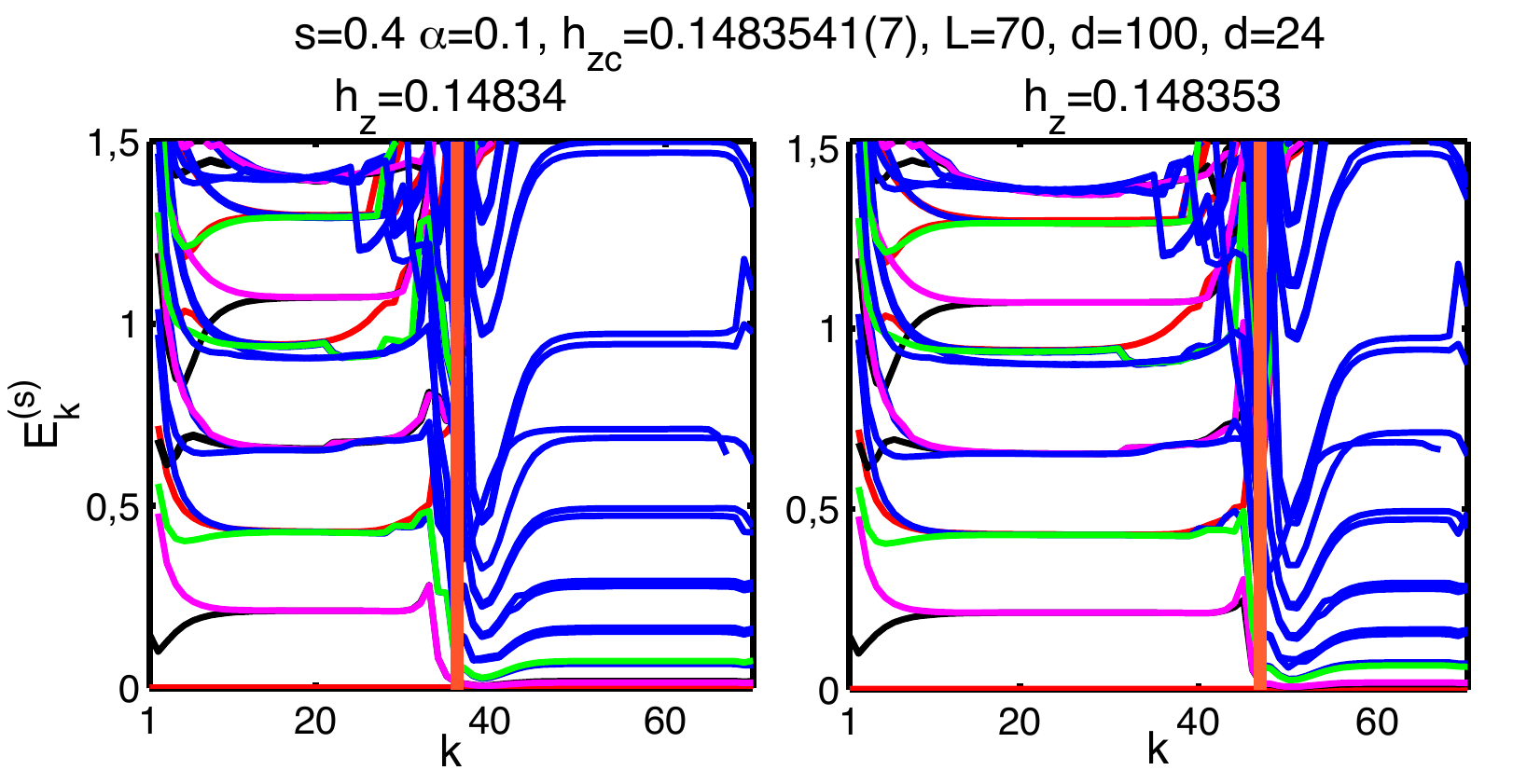}
\caption{Energy-level flow diagrams for $s=0.4$ in the LO close to the LO--DE transition. The smooth behavior in the first iterations reflect the characteristics of the critical fixed point while the bending and jumps in the lines suggest that the system flows to the localized fixed point. The red bar indicates the characteristic iteration $k^*$ of the transition that is used to calculate the low-energy scale $T^*$. By tuning $h_z$ close to its critical value, $k^*$ moves towards higher iterations.}
\label{fig:nu_flow}
\end{figure}

\subsubsection{Response exponent $\delta$}

As defined in \Eq{eq:delta}, $\delta$ can be extracted from the response at criticality of the order parameter $M_{xy}$ to an external magnetic field. \Fig{fig:exp_qc2}(b) displays the typical power-law scaling of the magnetization at the critical point. The deviations at small $h_x$ are again related to finite system size and numerical artifacts. Determining $\delta$ from power-law fitting over 6 decades for fixed $s=0.4$, we find it to be in accordance with the mean-field predictions of the quantum-to-classical mapping, $\delta_{\mathrm{MF}}=3$. Although the deep subohmic regime $s<0.3$ is again not accurately accessible for our VMPS approach, the collected results for  $s\geqslant0.3$ depicted in \Fig{fig:exp_qc2}(e) strongly  support the validity of quantum-to-classical mapping: for $s<1/2$ $\delta$ approaches its mean-field value of $\delta_{\mathrm{MF}}=3$, while for $1/2<s<1$ it clearly follows the hyperscaling relation in \Eq{hs5}.

\subsubsection{Correlation-length exponent $\nu$}
\label{sec:nuQC2}

The definition of the correlation-length exponents $\nu$ in \Eq{nu} involves a crossover energy scale $T^\ast$ that can easily be derived using the VMPS energy-flow diagrams introduced in Section~\ref{sec:flowdiag}. To this end, we determine the site on the Wilson chain $k^{\ast}$ where the flow starts to significantly deviate from the characteristically smooth flow at the critical point. This approach is illustrated in \Fig{fig:nu_flow}, where two typical energy flows inside the localized phase close to QC2 are displayed. In the beginning the system resides at the critical fixed point (smooth energy flow), then a transition to the localized fixed point occurs. This crossover is indicated by the red bar corresponding to the iteration $k^{\ast}$. It is defined by the point where the first excited energy level drops below $E<0.05$ in rescaled energy units.

\begin{figure}[t]
\centering
\includegraphics[width=.5\textwidth]{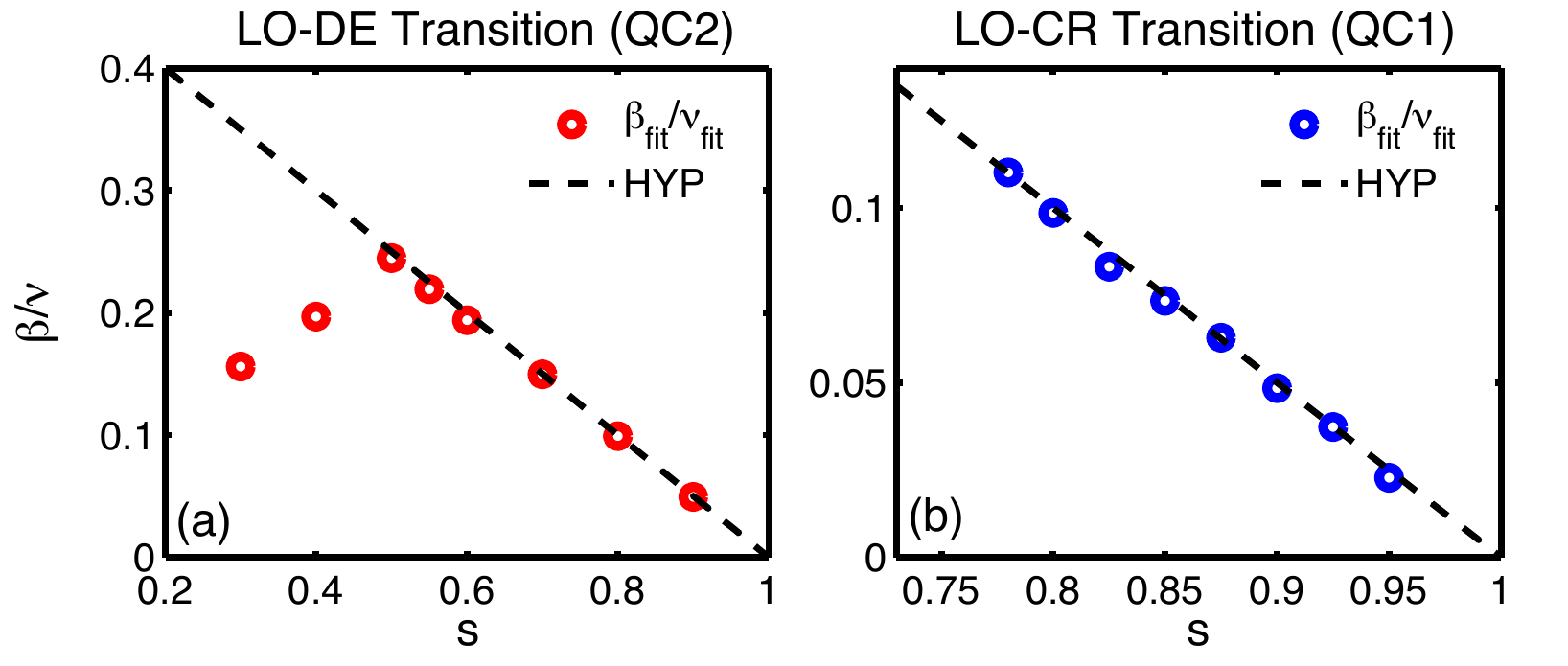}
\caption[]{Hyperscaling relation \eqref{hs4} involving exponents $\beta$ and $\nu$ at QC1 (a) and QC2 (b) with the numerical data (dots)  is compared to the exact results (dashed line). We find excellent agreement with the theory at QC1 for all values of $s^*<s<1$ as well as for $1/2<s<1$ at QC2. As expected by quantum-to-classical mapping, the numerical data confirm that hyperscaling (HYP) fails at QC2 below $s<1/2$. }
\label{fig:hyp}
\end{figure}

The crossover energy scales $T^\ast$ determined from such an analysis are collected in \Fig{fig:exp_qc2}(c) for fixed $s=0.4$ and $\alpha=0.1$ close to the phase transition. The power-law scaling of $T^\ast$ over several orders allows us to extract $\nu$ with high accuracy. Studying the $s$-dependence in \Fig{fig:exp_qc2}(f), we again find excellent agreement with the classical XY model:  $\nu$ closely follows the mean-field prediction [\Eq{eq:numf}] for $0<s<1/2$ (black dashed line), and also agrees with the perturbative RG calculations near $s=1/2$ [\Eq{eq:nu_rg}] and $s=1$ [\Eq{eq:nu_rg2}]. As a further check, we analyze the validity of the hyperscaling relation \Eq{hs4} involving both $\beta$ and $\nu$ by usage of our numerical results. \Fig{fig:hyp}(a) shows that the numerically extracted exponents obey hyperscaling for $1/2<s<1$ but clearly violate the respective relation in regime $s<1/2$, as expected by quantum-to-classical correspondence.

\subsection{Transition between LO and CR phases}

Next, we consider the second continuous quantum phase transition of {\twobm} between LO and CR phase, which is controlled by the critical fixed point QC1. In this case, the quantum-to-classical correspondence is presumably violated, see Section~\ref{sec:QCC}, and no analytical predictions for the critical exponents are available.

\subsubsection{Order parameter exponent $\beta$}

\begin{figure*}
\centering
\includegraphics[width=1\textwidth]{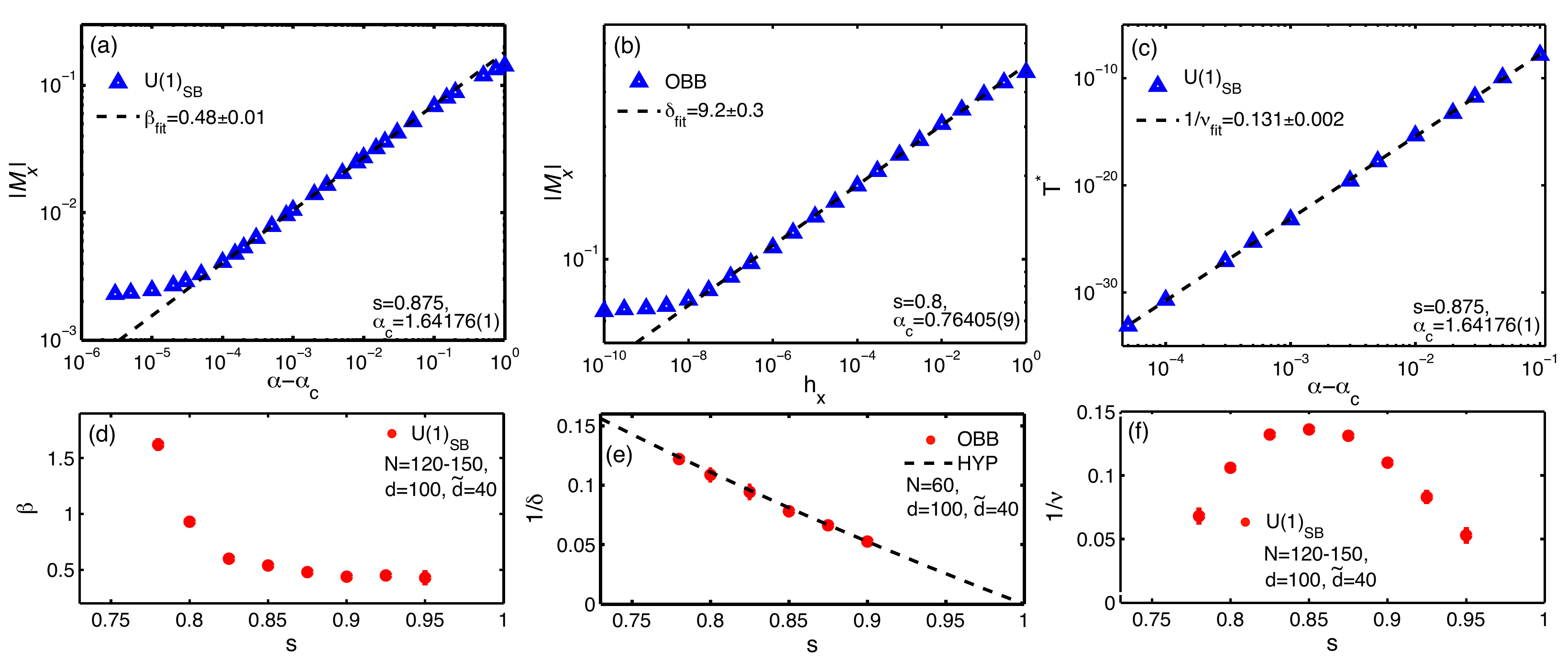}
\vspace{-15pt}
\caption{VMPS results for critical exponents $\beta$, $\delta$, and $\nu$ at the LO--CR quantum phase transition. In (a) the power-law scaling of the order parameter in the vicinity of the critical point is displayed for fixed $s=0.875$, whereas the fitted values of $\beta$ are collected for various $s$ in (d). Panel (b) shows similar VMPS data for the order parameter at the critical point w.r.t.\,an increasing $h_x$ for fixed $s=0.8$, which we use to extract the exponent $\delta$. Its overall $s$-dependence is illustrated in (e), which is in accordance with hyperscaling \Eq{hs5} (dashed line). In addition, (c) depicts the crossover energy scale $T^\ast$ close to the transition, which relates to the exponent $\nu$ for the correlation length which shows an overall $s$-dependence according to panel (f).}
\label{fig:exp_qc1}
\end{figure*}

The transition between CR and LO phase can be driven by varying $\alpha$ at $h_z=0$ and $s^*<s<1$. Hence, $\beta$ is defined according to Eq.~(\ref{eq:beta}) with the corresponding numerical data displayed in \Figs{fig:exp_qc1}(a) and (d). The upper panel (a) depicts the scaling of  $M_x$ close to $\alpha_c$ for fixed $s=0.875$ on a log-log scale, where a power-law behavior is apparent over more than 3 decades, with an exponent $\beta=0.48\pm0.01$. The lower panel (d) shows the dependence of $\beta$ on the bath exponent $s$ gained from power-law scaling fits with similar quality as  \Fig{fig:exp_qc1}(a). We find increasing values of $\beta>1$ for $s\to s^*$, while in the limit of $s\to 1^-$ our VMPS calculations suggest that $\beta$ approaches the value $1/2$.
Furthermore, we are able to show that $\beta$ in combination with $\nu$ satisfies the hyperscaling relation in \Eq{hs4}, as illustrated in \Fig{fig:hyp}(b).

Note that the extraction of $\beta$ is particularly complicated for QC1, since this transition comes with a large exponent $\nu$ for the correlation length, on which we elaborate below. This property relates to a low-energy scale required to resolve $\alpha_c$ appropriately \Eq{eq:accuarcy}, a precondition to obtain a solid power-law scaling of the order parameter $M_x$. Such calculations involving large chain lengths ($N>100$) become extremely sensitive to artificial symmetry breaking caused by numerical noise. Therefore, the use of the  symmetry-enforced VMPS is essential in this parameter regime, for performance and accuracy reasons. In particular, the application of OBB fails for energy scales significantly below double precision accuracy, since the small ``perturbations'' introduced by a shifted basis grow exponentially for later Wilson shells, and hence break the energy-scale separation on the Wilson chain. This should not affect the validity of our results, since a shifted basis is not strictly required for $1/2<s<1$ (see Appendix \ref{app1}).

\begin{figure}[b]
\centering
\includegraphics[width=.5\textwidth]{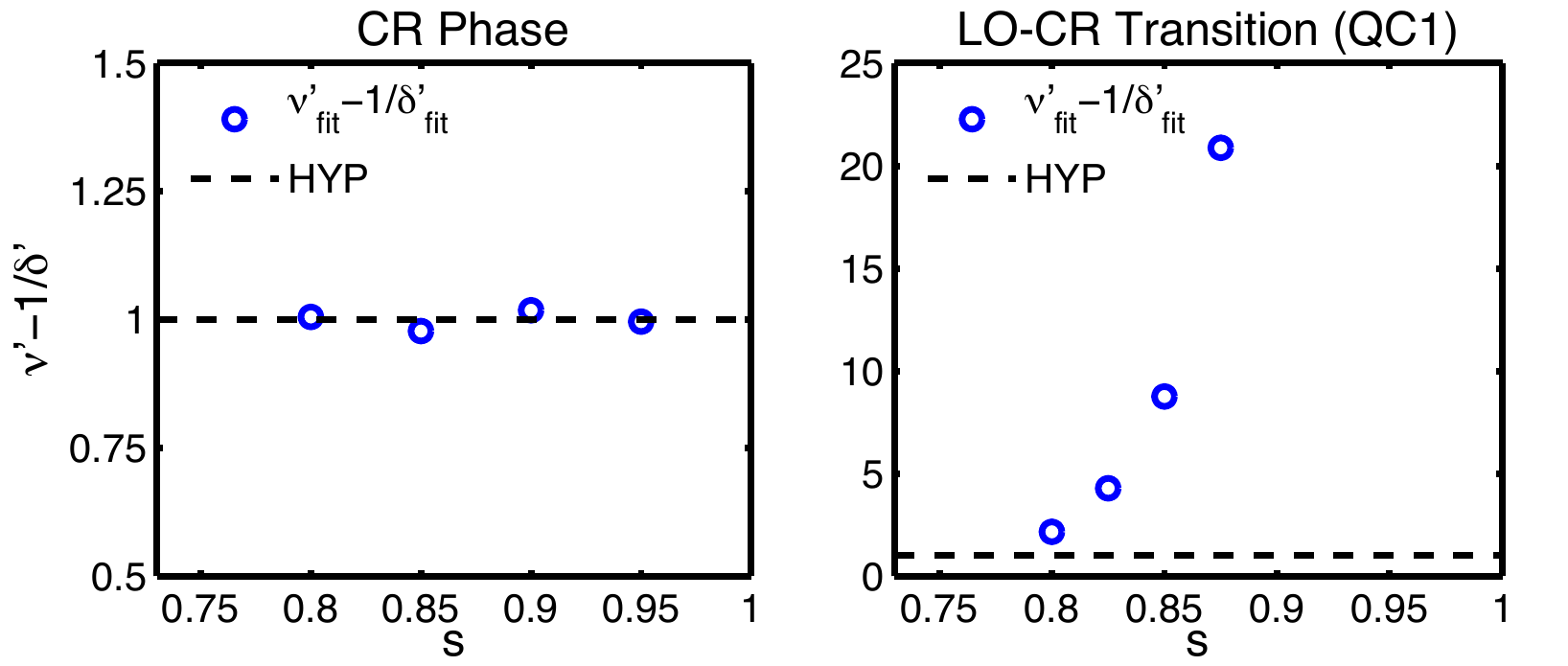}
\caption[]{Hyperscaling relation \eqref{eq:hyp_nup} involving exponents $\nu'$ and $\delta'$ in the CR phase (a) and at LO--CR critical point (b). The numerical results (dots) are in reasonable agreement with the hyperscaling relation (dashed line) in the CR phase (a). At QC1, however, hyperscaling appears to be violated. Note that the dots show $\nu'-1$, as we find that $\delta'=1$ for all $s$.}
\label{fig:hyp_nup}
\end{figure}

\subsubsection{Response exponents $\delta$ and $\delta'$}

For a transition at $h_z=0$, it is possible to extract both exponents $\delta$ and $\delta'$ via the order parameter's response to a magnetic field  according to \Eqs{eq:delta} and \eqref{deltapr}, respectively. Focusing first on $\delta$, \Figs{fig:exp_qc1}(b) and (e) illustrate the results of our VMPS calculations. Again, the upper panel (b) shows the typical response of the magnetization to an increasing $h_x$ at the critical point for $s=0.8$. The robust power-law scaling over more than six decades allows us to extract $\delta=9.2\pm0.3$ with high accuracy. The data collected from OBB calculations with different values of $s$ in the lower panel (e) indicates that $\delta$ closely follows the hyperscaling relation in \Eq{hs5}.

In contrast to $\delta$, we find the exponent $\delta'$, corresponding to the $h_z$ response, to be completely independent of the bath exponent $s$, having $\delta'=1$ for all $s$ at the LO--CR transition (not shown).

\subsubsection{Correlation-length exponent $\nu$}

As described above, the crossover energy scale $T^\ast$ characterizing the LO--CR transition is obtained by studying energy-flow diagrams close to the phase boundary. \Fig{fig:exp_qc1}(c) displays the extracted $T^\ast$ for fixed $s=0.875$ and $h_z=0$, with clear power-law scaling being apparent over several decades. This allows us to extract $\nu$ with high accuracy by fitting. \Fig{fig:exp_qc1}(f) shows the $s$ dependence of the exponent $\nu$. Our results suggest that $\nu$ diverges both in the limit $s\to s^{*+}$ and $s\to 1^{-}$, in a manner reminiscent of the approach to a lower critical dimension. We have verified that the exponent $\nu$ is identical for both sides of the transition, i.e., independent of whether QC1 is approached from the LO or from the CR phase.

Generally, the computed values of $\nu$ take large values for the entire range of bath exponents. As previously discussed, this causes our VMPS calculations to require large chains ($N>100$) in order to access the ultra-low energy scales needed to accurately determine $\alpha_c$.

\subsubsection{Correlation-length exponent $\nu'$, absence of hyperscaling, and field instability of QC1}

A finite $h_z$ applied at the zero-field critical coupling $\alpha_c(h_z\!=\!0)$ places the system into the DE phase. The characteristic crossover scale $T^{\ast}$ obtained from the energy-flow diagrams determines the critical exponent $\nu'$, which only diverges in the limit $s\to 1^{-}$ but not for $s\to s^{*+}$, in contrast to $\nu$. Most importantly, $\delta'$ and $\nu'$ in combination do not obey the hyperscaling relation \eqref{eq:hyp_nup}, as illustrated in \Fig{fig:hyp_nup}(b).
Hence, our results suggest that QC1 obeys hyperscaling properties only in the absence of a transverse field $h_z$. The underlying reason for this exotic critical behavior is not understood.

We note that the values for $\nu'$ can be read off from \Fig{fig:hyp_nup}(b) as $\delta'=1$ for all $s$. They imply that $\nu>\nu'$ for $s^\ast<s\lesssim0.83$ while $\nu<\nu'$ for $0.83\lesssim s < 1$, i.e., the role of the leading relevant operator at QC1 changes at $s\approx0.83$, see Fig.~\ref{fig:rgflow}(b).

\begin{figure}[t]
\centering
\includegraphics[width=0.5\textwidth]{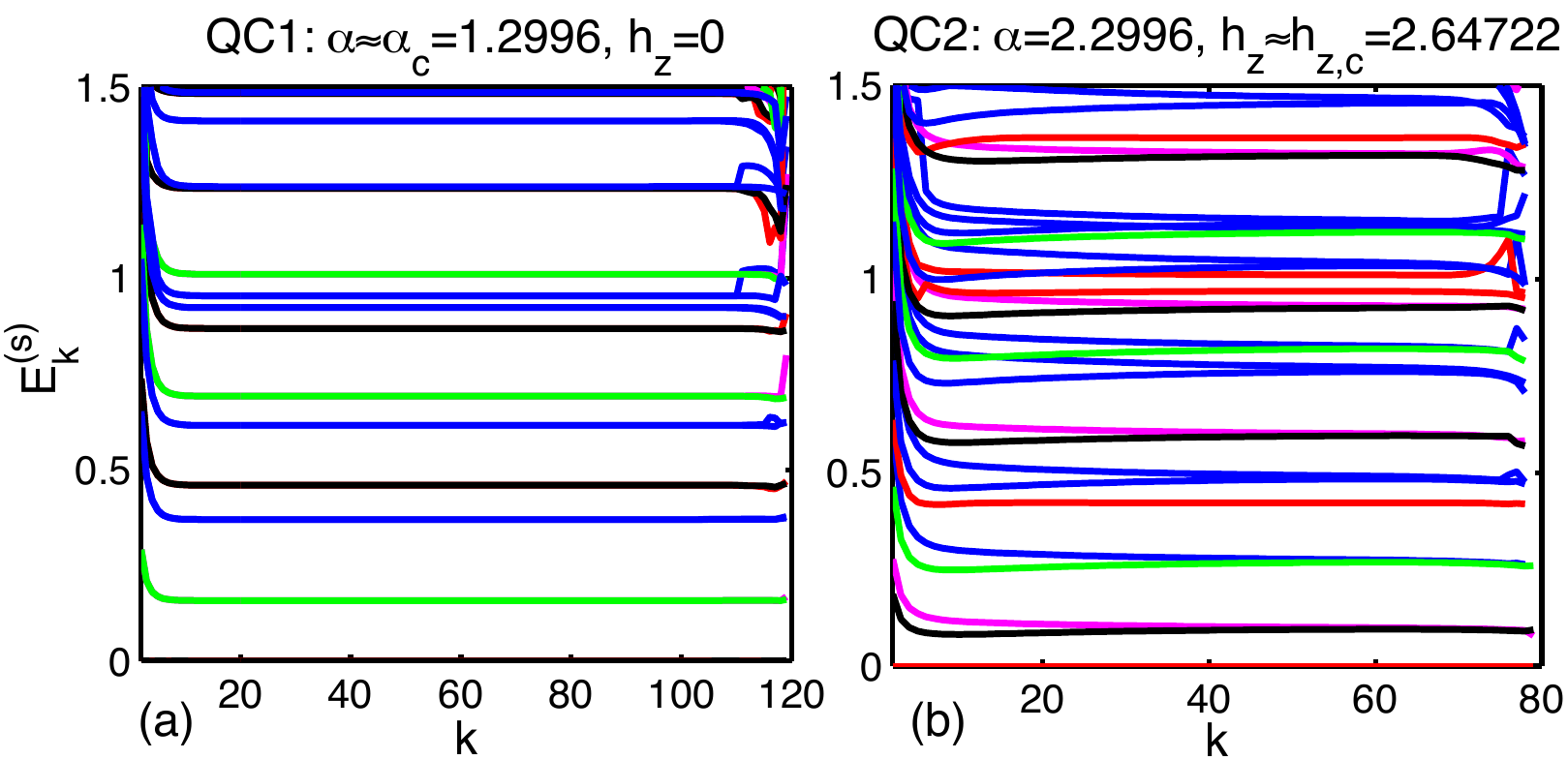}
\caption{Energy-flow diagrams at QC1 (a) and QC2 (b). The two critical fixed points can be distinguished by noting the two-fold ground state degeneracy at QC1 that disappears at QC2 introducing a finite $h_z$. }
\label{fig:QC1QC2_flow}
\end{figure}

\begin{figure}[t]
\centering
\includegraphics[width=0.5\textwidth]{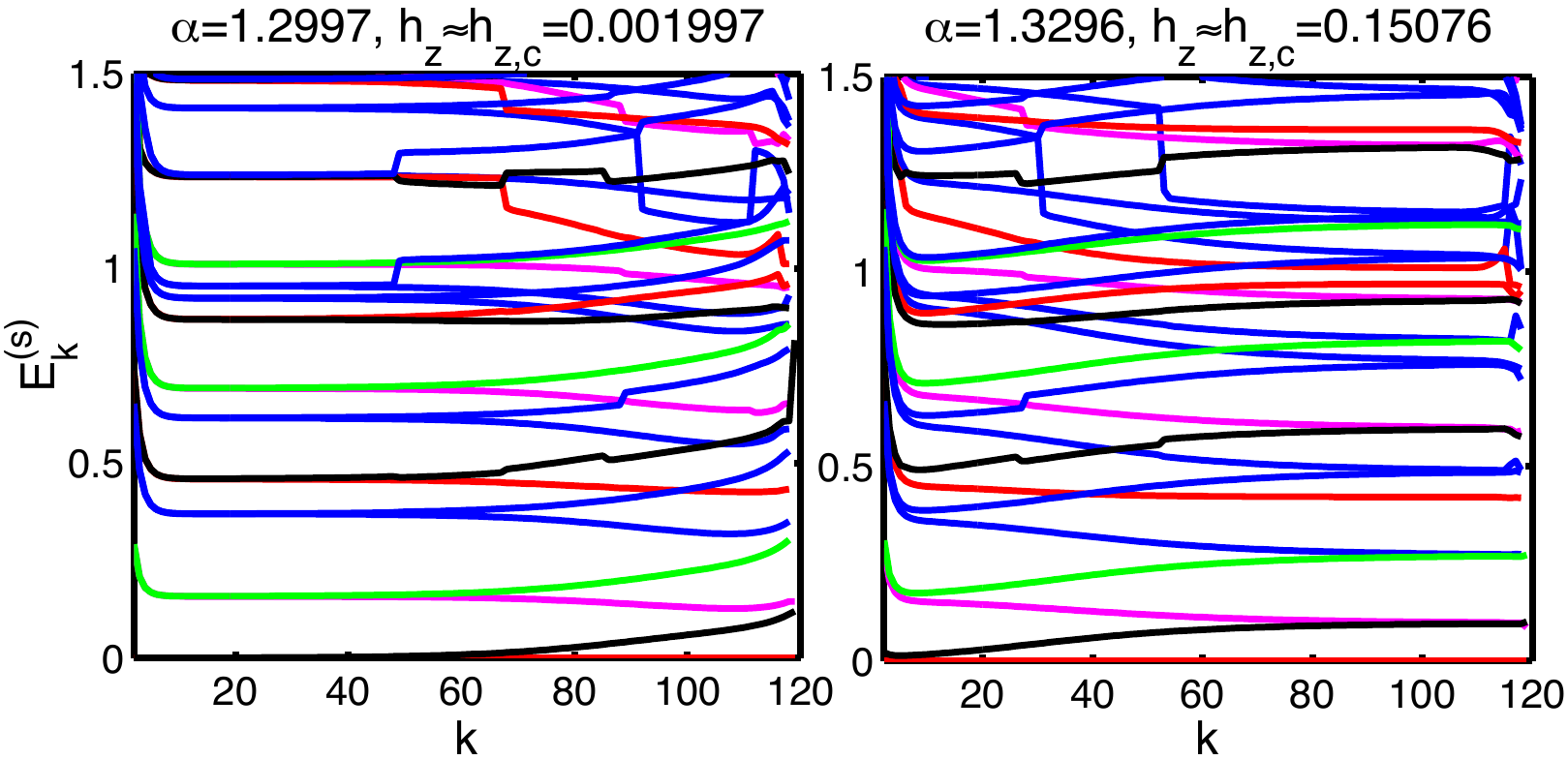}
\caption{Two energy-flow diagrams for parameters located on the critical separatrix with small $h_z$, i.e., close to QC1. In both cases, the level energies clearly flow from QC1 at high energies to QC2 at low energies, thus confirming the instability of QC1.}
\label{fig:QC1C2cross_flow}
\end{figure}

We have also investigated the flow {\em along} the separatrix between DE and LO at small $h_z$, in order to verify that QC1 is unstable along this separatrix, which implies that any finite-field transition is controlled by QC2. To this end, we first identify the stable energy-flow patterns corresponding QC1 and QC2 by placing the system at criticality for $h_z=0$ (QC1) and sizeable $h_z$ (QC2), see \Fig{fig:QC1QC2_flow}. Second, we study the energy-flow diagrams for parameters sets at criticality and very small $h_z$. As displayed in \Fig{fig:QC1C2cross_flow}, we observe a clear flow from QC1 at high energies to QC2 at lower energies, thus confirming the schematic RG flow diagram in Fig.~\ref{fig:rgflow}(b).

\subsection{CR phase}

We supplement the analysis of the critical phenomena of {\twobm} by briefly elaborating on the properties of the impurity spin in the CR phase. Although an abridged version of the results has already been presented elsewhere,\cite{guo_vmps_2012supp} this section completes the picture and also includes a discussion on the validity of hyperscaling inside the CR phase.

\subsubsection{Response exponents $\delta$ and $\delta'$}

\begin{figure*}
\centering
\includegraphics[width=1\textwidth]{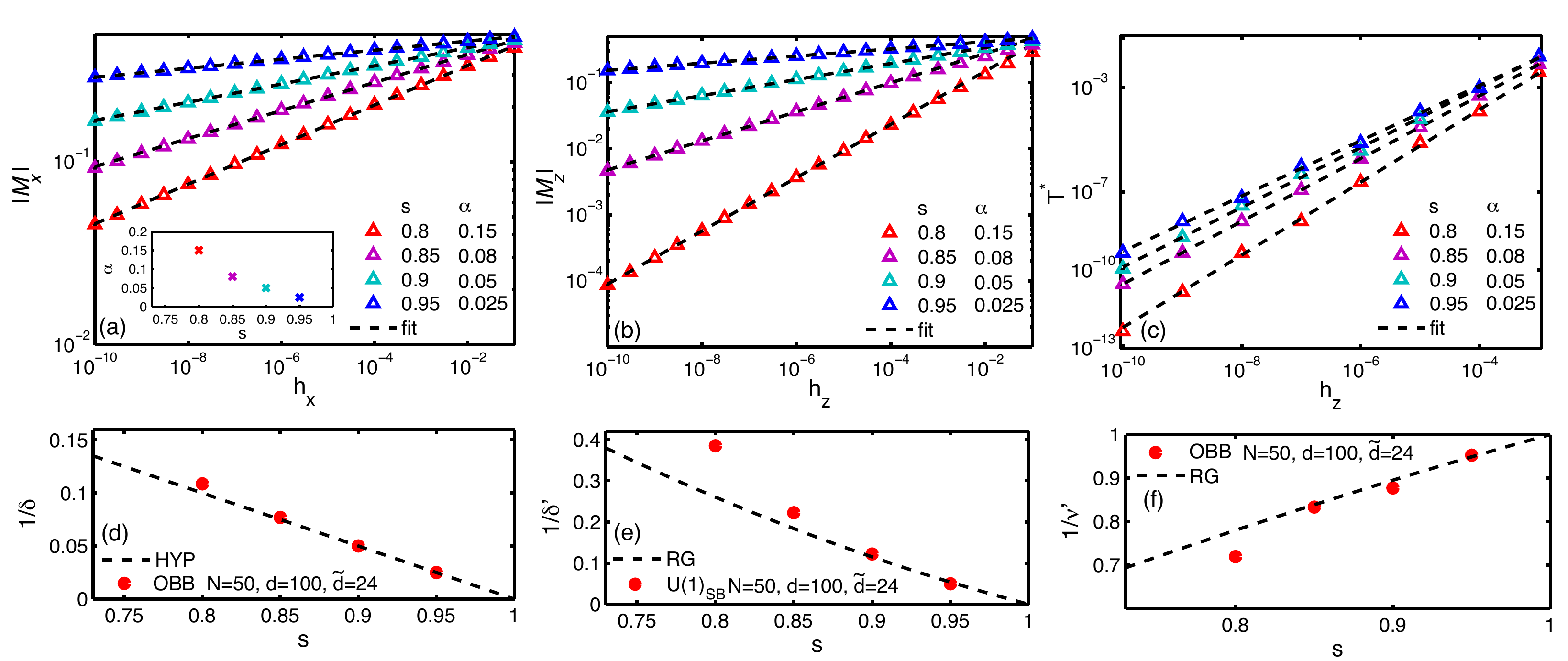}
\caption{
VMPS results for critical exponents $\delta$, $\delta'$ and $\nu'$ inside the CR phase for various $s$, with $\alpha$ chosen close to the CR fixed point $\alpha^\ast$ [$\alpha(s)$ displayed in inset of panel (a)]. The magnetization shows non-linear behavior in response to $h_x$ (a) and $h_z$ (b). The exponents $\delta$ and $\delta'$ extracted from the power-law scaling are in good agreement with perturbative RG  from \Eqs{eq:delta_cr} and \eqref{eq:deltap_cr} in the limit of $s\to 1^+$, as illustrated in panels (d) and (e). The same applies for $\nu'$ computed from the vanishing crossover energy scale $T^\ast$ in panel (c), which agrees with the RG prediction  \eqref{eq:nup_cr} for large values of the bath exponent $s$ (f).
}
\label{fig:exp_cr}
\end{figure*}

The RG calculations around the free-spin fixed point, presented in Section~\ref{sec:RGF}, predict a non-linear scaling of the magnetization in the CR phase, see \Eqs{eq:delta_cr} and \eqref{eq:deltap_cr}.
Our numerical data confirm this non-linear response, as illustrated in \Figs{fig:exp_cr}(a) for $\delta$ and in (b) for $\delta'$ at different values of $s$, $\alpha$ chosen close to the CR fixed point $\alpha^*$.
We find a clear power-law scaling over several decades. The extracted values for $\delta$ in Fig.~\ref{fig:exp_cr}(d) are perfectly consistent with the hyperscaling result \eqref{eq:delta_cr}, while those for $\delta'$ in Fig.~\ref{fig:exp_cr}(e) are in good agreement with the perturbative results for $s\to1^-$. The small deviations of the numerical data from the RG calculations for larger values of $(1-s)$ is expected, since the higher-order contributions in Eq.~\eqref{eq:deltap_cr} become more important.

\begin{figure}[b]
\centering
\includegraphics[width=.5\textwidth]{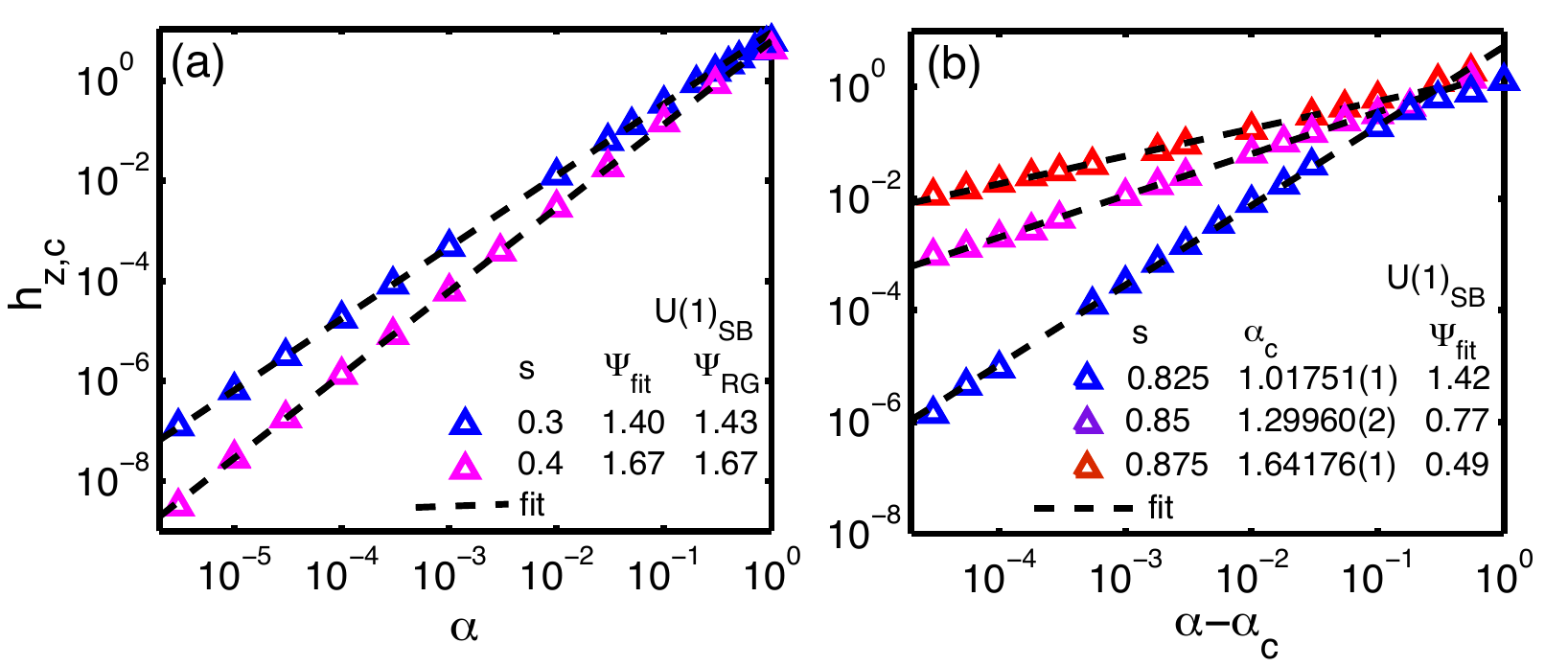}
\caption[]{Numerical results for the DE--LO phase boundary at small $h_z$ for selected values of the bath exponent $s$, obtained by U$(1)_{\rm SB}$ with $N=50, \tilde{d}_k=24$ (a) and $N=120, \tilde{d}_k=40$ (b).
(a) Regime of $s<s^\ast$ where $\hzc \propto \alpha^\psi$.
(b) Regime of $s^\ast<s<1$ where $\hzc \propto (\alpha-\alpha_c)^\psi$.
The power-law fits to determine the phase-boundary exponent $\psi$ are shown by dashed lines.
}
\label{fig:psi}
\end{figure}

\subsubsection{Correlation-length exponent $\nu'$ and field instability of CR}

The energy-flow diagrams (not shown) confirm that the CR phase is unstable w.r.t.\ a finite transverse field $h_z$, i.e., applying any finite $h_z$ places the system into the DE phase. The corresponding crossover scale $T^\ast(h_z)$ between the CR and DE fixed points allows us to extract the correlation-length exponent $\nu'$ \eqref{nupr}. The collected results for different $s$ are displayed in \Fig{fig:exp_cr}(c) and (f), with $\nu'$ being in fair agreement with perturbative prediction in Eq.~\eqref{eq:nup_cr}.

In contrast to QC1, where the hyperscaling relation  \eqref{eq:hyp_nup} is not met by the numerical data, $\nu'$ and $\delta'$ obey hyperscaling in the critical phase, as depicted in \Fig{fig:hyp_nup}(a).


\subsection{Phase-boundary exponent $\psi$}

We have determined the location of the DE--LO phase boundary for small $h_z$, in order to extract the expected power-law behavior. Sample results for $s<s^\ast$, where the phase boundary starts at $\alpha=0$, are shown in Fig.~\ref{fig:psi}(a); they are in essentially perfect agreement with the analytical result $\psi=1/(1-s)$ \eqref{psires1}.

For $s^\ast<s<1$ the phase boundary starts at the zero-field CR--LO transition at $\alpha=\alpha_c$, and thus determining $\psi$ requires an accurate knowledge of $\alpha_c$ and is therefore rather time-consuming. Sample results are in Fig.~\ref{fig:psi}(b). A hyperscaling-based guess would be $\psi=\nu/\nu'$ which we find approximately fulfilled for $s=0.825$ and $0.85$, but violated for $s=0.875$. (Recall that the hyperscaling relation between $\nu'$ and $\delta'$ is violated as well.)


\section{Conclusions}
\label{sec:concl}

Using the variational matrix-product-state approach, we have numerically determined the phase diagram of the U(1)-symmetric two-bath spin-boson model ({\twobm}), which is characterized by the phenomenon of frustration of decoherence.
Our detailed study of the quantum phase transitions of {\twobm}, using both numerical and analytical techniques, has revealed that the transition between the localized and delocalized phase, accessed at finite transverse field, is in the universality class of the XY spin chain with long-ranged interactions and thus obeys a quantum-to-classical correspondence.

In contrast, the zero-field critical (intermediate-coupling) phase and its transition to the localized phase do not have a classical counterpart. Our numerical results for the critical exponents can serve as a guide for developing an analytical theory of the latter transition. Given that the relevant critical fixed point (QC1) approaches the localized fixed point (LO) as $s\to1^-$, we believe that an expansion around LO akin to an expansion in $(2+\epsilon)$ dimensions for classical magnets should be able to access the properties of QC1 -- this task is left for future work.

We recall that the analysis in this paper has been restricted to the model {\twobm} with symmetric couplings, i.e., two identical baths and $\alpha_x = \alpha_y$. For asymmetric couplings, with finite $\Delta \alpha = \alpha_y -\alpha_x$, the behavior of the model is driven towards that of the one-bath model {\onebm}. Naturally, the LO phase now displays spontaneous Ising order, with the impurity spin localized in direction of the stronger coupled bath. Further, the CR phase is unstable against any finite $\Delta\alpha$.
The rich and interesting crossover physics of {\twobm} in the presence of small symmetry breaking is beyond the scope of this paper and will be discussed elsewhere.

Interesting open questions concern the finite-temperature behavior of {\twobm}, specifically the quantum critical finite-$T$ susceptibilities and the residual entropy, as well as its equilibrium and non-equilibrium dynamics. Generalizations to three bosonic baths as well as combined fermionic and bosonic baths would be interesting as well. The former is linked to the problem of impurity spins in quantum critical magnets,\cite{anirvan,VBS} and both occur in self-consistent single-site solutions for certain lattice models, e.g., in the large-$N$-based theory of a gapless spin liquid \cite{ssye} and in more general extensions of dynamical mean-field theory.\cite{edmft}

In the quest for non-trivial quantum critical behavior, we believe that {\twobm} presents -- in a sense -- the simplest quantum model violating the quantum-to-classical correspondence: It lives in $(0+1)$ dimensions and is constructed solely from bosonic degrees of freedom. Our analysis reveals that the violation of the quantum-to-classical correspondence is rooted in the non-commutativity of the spin components coupled to the two baths; this property of a quantum spin can also be re-phrased as a spin Berry phase.
We note that quantum phase transitions in quantum impurity models with {\em fermionic} baths frequently behave non-classically, with the pseudogap Kondo and Anderson models\cite{pgkondo1,pgkondo2} being well-studied examples. Here, the absence of a quantum-to-classical correspondence can be traced back to fermionic ``signs'', i.e., exactly integrating out the fermionic bath is only possible at the expense of working with a fermionic impurity, which has no classical analogue.


\acknowledgments

We thank A. Alvermann, S. Florens, S. Kirchner, K. Ingersent, Q. Si, A. Schiller,
and T. Vojta for helpful discussions.  This research was supported by
the Deutsche Forschungsgemeinschaft through the Excellence Cluster "Nanosystems
Initiative Munich", SFB/TR~12, SFB~631, WE4819/1-1 (AW), FOR~960, by
the German-Israeli Foundation through G-1035-36.14, and the NSF through
PHY05-51164.


\appendix

\section{Scaling hypothesis for QC1}
\label{app:hyper}

Here we sketch the use of the scaling hypothesis to deduce hyperscaling relations for QC1. The standard homogeneity law for the critical contribution to the free energy implies the scaling form
\begin{equation}
\label{scalhyp}
F_{cr}(\alpha,h_x,h_z,T) = T f_1\left(\Delta\alpha/T^a, h_x/T^b, h_z/T^c\right),
\end{equation}
recall that the problem at hand is effectively (0+1)-dimensional. Here, $\Delta\alpha=\alpha-\alpha_c$ and $h_z$ correspond to two operators which drive the system away from criticality, and $f_1$ is a scaling function. The definitions of the correlation-length exponents in Eqs.~\eqref{nu} and \eqref{nupr} lead to the identifications $a=1/\nu$ and $c=1/\nu'$.

Taking the derivative of Eq.~\eqref{scalhyp} w.r.t. $h_x$ yields
\begin{equation}
M_x = T^{1-b} f_2\left(\Delta\alpha/T^a, h_x/T^b, h_z/T^c\right)
\end{equation}
which can be cast into the forms
\begin{equation}
\label{mx1}
M_x = (\Delta\alpha)^{(1-b)/a} f_3\left(T^a/\Delta\alpha, T^b/h_x, T^c/h_z\right)
\end{equation}
and
\begin{equation}
\label{mx2}
M_x = h_x^{(1-b)/b} f_4\left(T^a/\Delta\alpha, T^b/h_x, T^c/h_z\right)\,.
\end{equation}
Upon taking the limit $T\to0$ in Eq.~\eqref{mx1} one deduces the order-parameter exponent as $\beta=(1-b)/a$; similarly Eq.~\eqref{mx2} yields $1/\delta = (1-b)/b$ or $b=\delta/(1+\delta)$. Using $a=1/\nu$ then leads to $\beta=\nu/(1+\delta)$ which is consistent with the hyperscaling relations \eqref{hs4} and \eqref{hs5}. Taking the second derivative of Eq.~\eqref{scalhyp} w.r.t. $h_x$ yields $\chi_x$ and facilitates the identification $1-2b = -x$. Together with $b=\delta/(1+\delta)$ this yields $\delta=(1+x)/(1-x)$, consistent with the relations \eqref{hs3} and \eqref{hs5}.

In full analogy, taking the derivative of Eq.~\eqref{scalhyp} w.r.t. $h_z$ yields $1/\delta' = (1-c)/c$ or $c=\delta'/(1+\delta')$. Using $c=1/\nu'$ finally gives $\nu' = 1/\delta' + 1$ which is Eq.~\eqref{eq:hyp_nup}. Taking the second derivative w.r.t. $h_z$ yields $1-2c= -x'$ and then $\delta'=(1+x')/(1-x')$ which is Eq.~\eqref{hs6}. The hyperscaling relations \eqref{eq:hyp_nup} and \eqref{hs6} can also be applied in the CR phase where $h_z$ corresponds to a relevant operator as well.
Note that the nature of the exponent pair $(\nu,\delta)$ is different from that of $(\nu',\delta')$: $\nu$ parameterizes the scaling dimension of $\Delta\alpha$ at criticality and $\delta$ the non-linear response to a field conjugate to the order parameter. In contrast, $\nu'$ and $\delta'$ correspond to the scaling dimension of and the non-linear response to the {\em same} field, $h_z$.


\section{Finite-size effects}
\label{app:finite}

As the numerical computations are done for finite Wilson chains, it is worth discussing finite-size effects arising from a finite chain length $N$. The most important effect of finite $N$ is to induce a gap $\bar\Delta$ in the bath spectrum which scales as $\bar\Delta \propto \Lambda^{-N}$. While this gap has no effect in the DE phase, as its fixed point corresponds to $\alpha=0$, it prevents true spontaneous symmetry breaking in the LO phase. However, this does not affect our calculations because, with increasing $N$, the finite correlation length induced by $\bar\Delta$ increases faster than the system size, such that the finite-size system ``looks'' ordered in the LO phase once $N$ is sufficiently large.

\begin{figure}
\centering
\includegraphics[width=.5\textwidth]{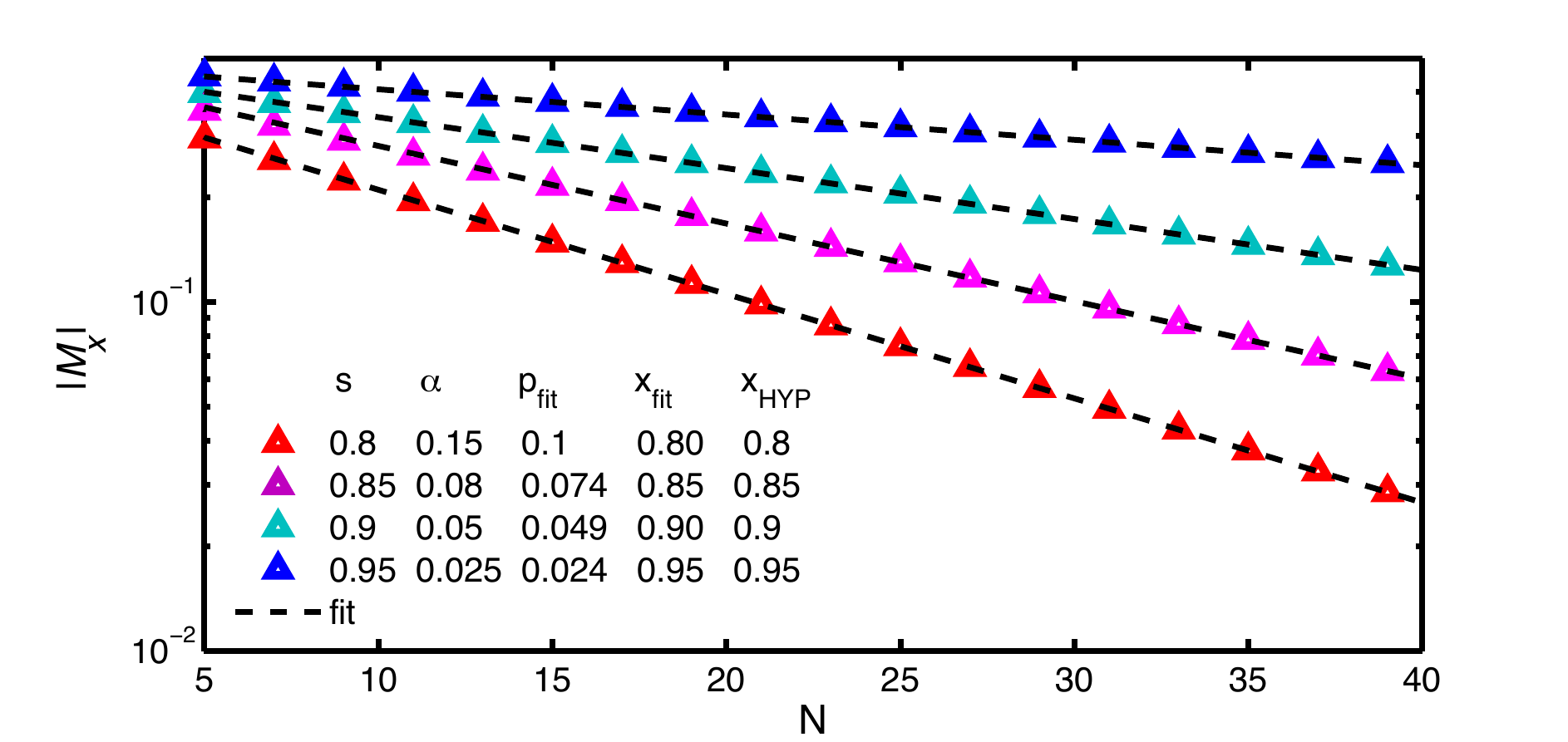}
\caption{Finite-size scaling of $M_x$ for different points close to the CR fixed point $\alpha^*$. We observe that the magnetization decreases exponentially with system size. The small value of the decay exponent $p$ results in a notable finite-size effects even for large systems [U$(1)_{\rm SB}$ with $N=60, \tilde{d}_k=24$].}
\label{fig:app_CRfss}
\end{figure}

Most problematic are finite-size effects in the CR phase. Here $M=0$ in the infinite-system limit, but a bath gap induces a finite residual magnetic moment scaling as\cite{VBS} $M \propto \bar{\Delta}^{(1-x)/2}$, with $x$ defined in Eq.~\eqref{xdef}.
Indeed, our VMPS calculations in \Fig{fig:CRmag} find a small but finite magnetization.  \Fig{fig:app_CRfss} supports that $M_x$ indeed vanishes in infinite-system limit as
\begin{eqnarray}
M_x \propto  L^{-p}  \propto (\Lambda^{-N})^p  =  e^{-\ln{(\Lambda)} p N}\,, \label{eq:fss}
\end{eqnarray}
with the system size $L \sim \Lambda^{N}$ on a Wilson chain. Given that the exponent $p$ governing the decrease of $M_x$ is very small, the order parameter remains finite even for very large systems.

The fit exponent $p$ allows us to extract the value of the exponent $x$ in the CR phase according to $p=(1-x)/2$. The values for $x$ obtained in this way are indicated in \Fig{fig:app_CRfss} and are consistent with the hyperscaling result $x=s$, see \Sec{sec:scaling}. We note that a direct measurement of $x$ at the various critical points is not easily possible using the present numerics, as (i) the variational approach is designed for $T=0$ only, and (ii) the mass-flow problem\cite{flow} would prevent an accurate approach to critical points using chains of different length.


\section{Influence of truncation error on critical exponents}
\label{app1}
As numerical artifacts play an increasingly important role close to the critical phase, we found it to be essential to enforce the conservation of the U(1) symmetry when trying to access the critical properties of QC1 and QC2 for  large bath exponents $s>0.8$. Since the symmetry incorporation excludes employing a shifted OBB-VMPS calculation, it is fair to ask whether the bosonic truncation error corrupts the presented results of the critical exponent $\beta$.

\begin{figure}[h!]
\centering
\includegraphics[width=.5\textwidth]{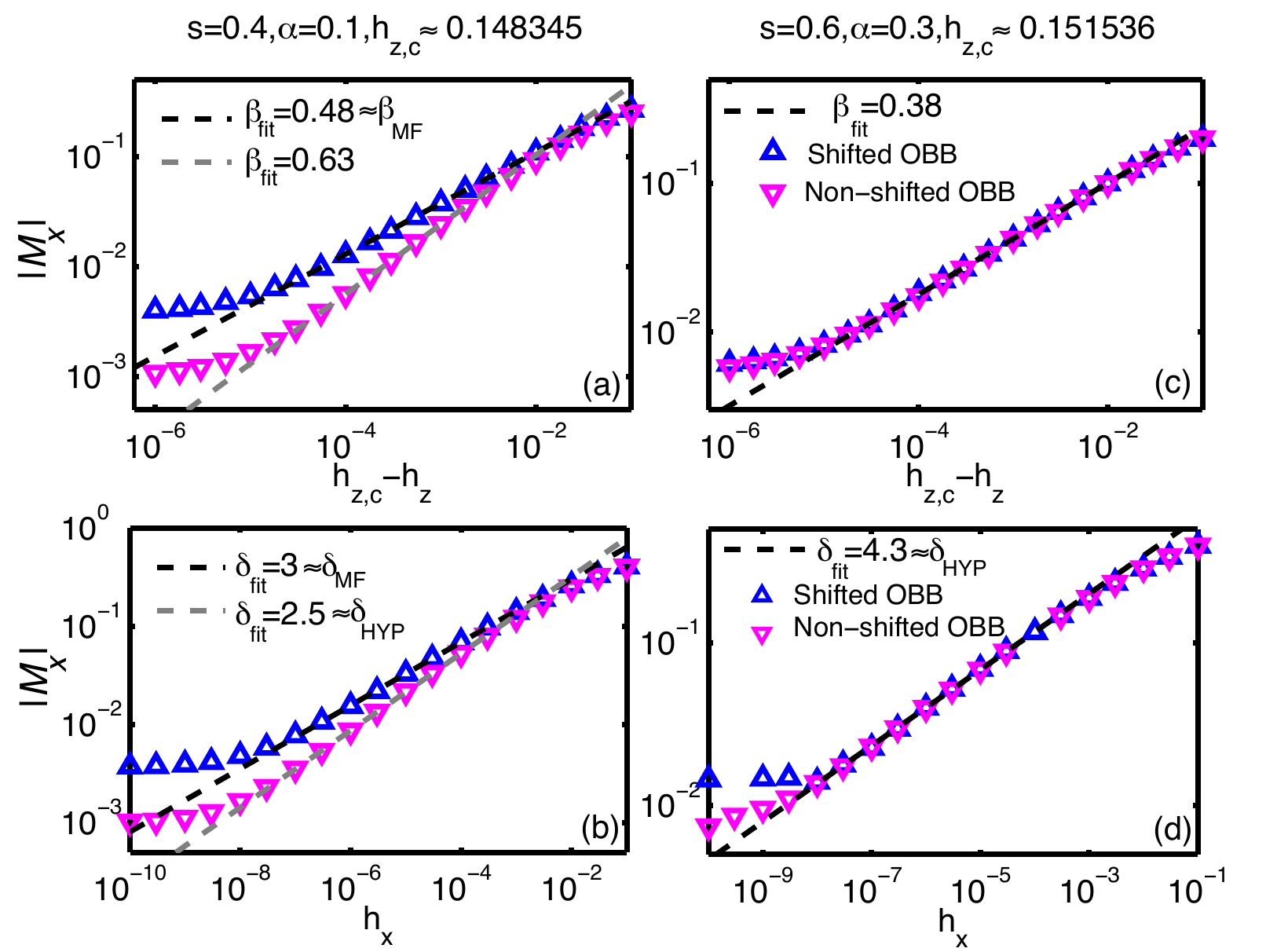}
\caption[]{Influence of Hilbert-space truncation on critical exponents $\beta$ and $\delta$ employing VMPS with (blue) and without shifted OBB (purple). Choosing $s=0.4<1/2$ in panels (a) and (b), we observe considerable deviations between both types of VMPS calculations where the mean-field prediction for $\beta$ and $\delta$ is only obtained with a shifted OBB. In the case of $s>1/2$, both methods lead to similar results, as illustrated in panels (c) and (d) for the exponent $s=0.6$ [OBB with $N=60, \tilde{d}_k=24$]. }
\label{fig:app_exp}
\end{figure}

Careful analysis revealed a similar situation as in the {\onebm},\cite{vojta_NRG_2012} where the resulting critical exponents are only affected by the truncation error in the regime $s<1/2$. In the same fashion, Hilbert-space truncation in the {\twobm} only influences the behavior of critical properties for $s<1/2$, as illustrated in \Fig{fig:app_exp}. Comparing the scaling of the magnetization to determine $\beta$ and $\delta$ using VMPS calculations with and without shifted OBB reveals the characteristic difference between $s<1/2$ and $s>1/2$. In \Fig{fig:app_exp}(a) and (b) we observe significant deviations between both types of calculations for $s=0.4$. Employing the shifted OBB method, the resulting critical exponents are in good agreement with the mean-field predictions $\beta_{\mathrm{MF}}=1/2$ and $\delta_{\mathrm{MF}}=3$, while VMPS calculations without shift lead to considerable deviations from the mean-field values. In contrast, considering a larger bath exponent $s=0.6$, we clearly obtain the same results for both types of VMPS calculations, as illustrated in \Fig{fig:app_exp}(c) and (d). Thus for the evaluation of critical exponents we conclude: the shifted OBB is only strictly necessary for small bath exponents $s<1/2$, whereas for of $s>1/2$ VMPS calculations with and without shifted OBB work equally well.

\section{Calculation of the magnetization in the U(1)-symmetric implementation}
\label{app2}

The ground state of {\twobm} in the LO phase with $h_x=h_y=0$ exhibits a continuous degeneracy due to the inherent rotational symmetry, which was elaborated on in Section~\ref{sec:U1}. When not enforcing the U(1) symmetry, the final ground state of a VMPS calculation spontaneously breaks this U(1) symmetry, while maximizing magnetization $M_x=M_y$ in $x$- and $y$-direction in the localized phase (note that this is the least entangled state).

In contrast, for a U(1)-symmetric implementation, these expectation values vanish by construction. However, it is possible to attach a well-defined symmetry label ($q=\pm 1/2$) to the numerical ground state. The two resulting states $|G_{\pm1/2} \rangle$ form an orthonormal pair, which can be used to construct the space of all (symmetry-broken) ground states.
By symmetry, the expectation value $\langle G_q | \sigma_{i=x,y} | G_{q'} \rangle$ evaluated using only one symmetry eigenstate, $q=q'$, gives zero. To reconstruct the magnetization of the ``original'', symmetry-broken ground state, we have to calculate the magnetization using non-diagonal elements  $q\neq q'$ of the above defined expectation value.

In general, this can be accomplished in two different ways. The simple but numerically expensive variant is to use two VMPS runs to obtain $|G_{+1/2}\rangle$ and $|G_{-1/2}\rangle$  separately for the same parameters and explicitly calculated $\langle G_{+1/2} | \sigma_{i=x,y} | G_{-1/2}\rangle$. Alternatively, we may borrow a concept of NRG that allows us to use only a single VMPS to determine the magnetization for a system with arbitrary Wilson chain length $0< k < N$. Starting with the right-orthogonalized representation of either $|G_{+1/2}\rangle$ or $|G_{-1/2}\rangle$, we construct and diagonalize the left block Hamiltonian $\HM^{k}_L$. After projecting into the subspace of the two lowest-lying energy states $|s_k\rangle$, with $s_k\in\{0,1\}$,
\begin{equation}
|s_k\rangle = \sum_{n_1 ... n_n} \big( A^{[\sigma]} A^{[n_1]} \,..\,A^{[n_{n}]} \big)_s |n_1,n_2,\,...\,,n_{n}\rangle\,,
\end{equation}
we explicitly determine all matrix elements $(M_i)_{s_ks_k'}^{[n]} \equiv  \langle s_k|\sigma_{i=x,y} |s_k'\rangle_n$ of the magnetization. The eigenvalues of the $2 \times 2$ matrix $M_i^{[n]}$  give the two possible values of the magnetization of the system with chain length $k$ in the ground state $\langle \sigma_{i=x,y}\rangle /2 = \pm M_i$.  Therefore the plain thermal average \emph{without spontaneous symmetry breaking} would result in zero magnetization.

\begin{figure}
\centering
\includegraphics[width=.51\textwidth]{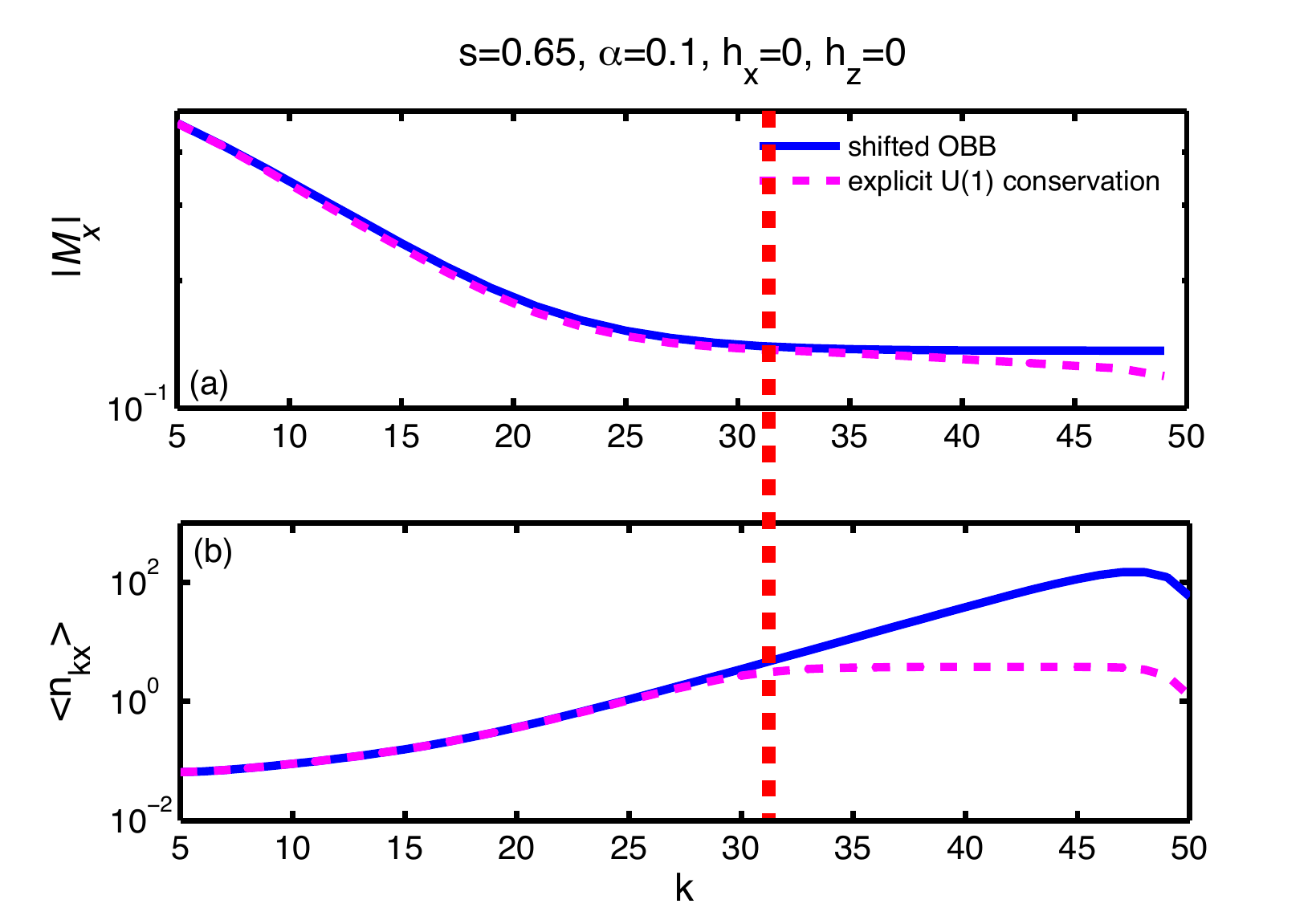}
\caption[]{Influence of bosonic truncation error. Studying the finite-size scaling effects of the impurity magnetization in the localized phase, we clearly observe in (a) that the symmetry implementation is accompanied by a further down-bending induced by reaching the maximum bosonic occuption numbers towards larger iterations.  }
\label{fig:app_fss}
\end{figure}

Independently of how the magnetization is calculated, we face an additional challenge regarding the Hilbert space truncation error  in the context of explicit symmetry implementation. Studying the finite-size scaling of the magnetization in the localized regime, we expect $M_x$ to saturate at a finite value after an initial decay when moving towards larger systems. As illustrated in \Fig{fig:app_fss}(a), our VMPS data for $M_x$ indeed saturates as expected when symmetry is not enforced (solid line).

However, when employing the symmetry implementation we observe a further  decrease (dashed line) after the saturation to an intermediate plateau. Considering the behavior of the bosonic occupation numbers on the Wilson chain in \Fig{fig:app_fss}(b), we attribute this effect with the Hilbert-space truncation error. The difference between solid and dashed lines sets in once the bosonic occupation numbers $\langle n_{kx} \rangle$ for the symmetry-enforced implementation, where a shifted OBB cannot be used, begin to saturate (\Fig{fig:app_fss}(b), dashed line), whereas those for non-symmetry-enforced implementation, for which a shifted OBB can be used, do not yet saturate (\Fig{fig:app_fss}(b), solid line). To circumvent this systematic error, we extract  $M_x$ not at the end of the chain but choose an iteration $N^*$ right before $\langle n_{k x}  \rangle$ saturates (indicated by the red dashed line in \Fig{fig:app_fss}). At $N^*$ the magnetization from the symmetry-enforced code clearly agrees with a VMPS calculation using a shifted bosonic basis.

As indicated in the previous section, this approach is only appropriate for bath exponents $s>1/2$. For smaller values of $s$ it is absolutely necessary to employ VMPS with the shifted OBB scheme in order to capture the correct physical properties of the system.



\begin{thebibliography}{99}

\bibitem{hewson} A. C. Hewson, \textit{The Kondo Problem to Heavy Fermions} (Cambridge University
  Press, New York, 1993).

\bibitem{NB} P. Nozi{\`e}res and A. Blandin,
J. Phys. (Paris) {\bf 41}, 193 (1980).

\bibitem{mvrev}
M. Vojta, Phil. Mag. {\bf 86}, 1807 (2006).

\bibitem{logan14}
D. E. Logan, A. P. Tucker, and M. R. Galpin,
Phys. Rev. B {\bf 90}, 075150 (2014).

\bibitem{rosch03}
A. Rosch, J. Paaske, J. Kroha, and P. W\"olfle,
Phys. Rev. Lett. {\bf 90}, 076804 (2003).

\bibitem{wilson_rev_1975}
K.~Wilson,
\rmp {\bf 47} 773 (1975).

\bibitem{bulla_rev_2008}
R.~Bulla, T.~Costi, and T.~Pruschke,
\rmp {\bf 80}, 395 (2008).

\bibitem{haas30}
W. J. de Haas, J. de Boer, and G. J. van den Berg,
Physica (Amsterdam) {\bf 1}, 1115 (1934).

\bibitem{rosch08}
T. A. Costi, L. Bergqvist, A. Weichselbaum, J. von Delft, T. Micklitz, A. Rosch, P. Mavropoulos, P. H. Dederichs, F. Mallet, L. Saminadayar, and C. B\"auerle,
Phys. Rev. Lett. {\bf 102}, 056802 (2009) .

\bibitem{cronen98}
S. M. Cronenwett, T. H. Oosterkamp, and L. P. Kouwenhoven,
Science {\bf 281}, 540 (1998).

\bibitem{gg98}
D. Goldhaber-Gordon, H. Shtrikman, D. Mahalu, D. Abusch-Magder, U. Meirav, and M. A. Kastner,
Nature {\bf 391}, 156 (1998).

\bibitem{makhlin}
Y. Makhlin, G. Sch\"on, and A. Shnirman,
Rev. Mod. Phys. {\bf 73}, 357 (2001).

\bibitem{garg85}
A. Garg, J. N. Onuchic, and V. Ambegaokar,
J. Chem. Phys. {\bf 83}, 4491 (1985).

\bibitem{demler13}
J. Bauer, C. Salomon, and E. Demler,
Phys. Rev. Lett. {\bf 111}, 215304 (2013).

\bibitem{nishida13}
Y. Nishida,
Phys. Rev. Lett. {\bf 111}, 135301 (2013).


\bibitem{leggett}
A.~J. Leggett, S. Chakravarty, A.T. Dorsey, M.P.A. Fisher, A. Garg, and W. Zwerger,
Rev. Mod. Phys. {\bf 59}, 1 (1987).

\bibitem{kehrein_spin-boson_1996}
S.~K. Kehrein and A.~Mielke,
Phys. Rev. A {\bf 219}, 313 (1996).

\bibitem{BTV03}
R.~Bulla, N.~Tong, and M.~Vojta,
Phys. Rev. Lett. {\bf 91}, 170601 (2003).

\bibitem{VTB}
M. Vojta, N. Tong, and R. Bulla,
Phys. Rev. Lett. {\bf 94}, 070604 (2005).

\bibitem{fehske}
A. Alvermann and H. Fehske,
Phys. Rev. Lett. {\bf 102}, 150601 (2009).

\bibitem{winter09}
A.~Winter, H.~Rieger, M.~Vojta, and R.~Bulla,
Phys. Rev. Lett. {\bf 102}, 030601 (2009).

\bibitem{VTB_err}
M.~Vojta, N.~Tong, and R.~Bulla,
Phys. Rev. Lett. {\bf 102}, 249904(E) (2009).

\bibitem{guo_vmps_2012}
C.~Guo, A.~Weichselbaum, J.~von Delft, and M.~Vojta,
\prl {\bf 108}, 160401 (2012).

\bibitem{vojta_NRG_2012}
M.~Vojta,
Phys. Rev. B {\bf 85}, 115113 (2012).

\bibitem{flow}
M. Voj\-ta, R. Bulla, F. G\"uttge, and F. B. Anders,
\prb {\bf 81}, 075122 (2010).

\bibitem{fisher_critical_1972}
M.~E. Fisher, S.~K. Ma, and B.~G. Nickel,
Phys. Rev. Lett. {\bf 29}, 917 (1972).

\bibitem{koster} J. M. Kosterlitz, \prl {\bf 37}, 1577 (1976).

\bibitem{luijten_classical_1997}
E.~Luijten and H.~W.~J. Bl\"{o}te,
Phys. Rev. B {\bf 56}, 8945 (1997).

\bibitem{anirvan} A.~M.~Sengupta,
\prb {\bf 61}, 4041 (2000).

\bibitem{rg_bfk}
L. Zhu and Q. Si, Phys. Rev. B {\bf 66}, 024426 (2002);
G. Zarand and E. Demler, Phys. Rev. B {\bf 66}, 024427 (2002).

\bibitem{antonio_xy}
A. H. Castro Neto, E. Novais, L. Borda, G. Zarand, and I. Affleck, Phys. Rev. Lett. {\bf 91}, 096401 (2003);
E. Novais, A. H. Castro Neto, L. Borda, I. Affleck, and G. Zarand, Phys. Rev. B  {\bf 72}, 014417 (2005).

\bibitem{dima}
D. V. Khveshchenko, Phys. Rev. B {\bf 69}, 153311 (2004).

\bibitem{sushkov98}
V. N. Kotov, J. Oitmaa, and O. Sushkov,
Phys. Rev. B {\bf 58}, 8500 (1998).

\bibitem{VBS}
M.~Vojta, C.~Buragohain and S.~Sachdev,
Phys. Rev. B {\bf 61}, 15152 (2000).

\bibitem{andrei1984}
N.~Andrei and C.~Destri, Phys.~Rev.~Lett.~{\bf 52}, 364 (1984).

\bibitem{wiegmann1985}
P.~B.~Wiegmann and A.~M.~Tsvelik, Z.~Phys.~B {\bf 54}, 201 (1985).

\bibitem{guo_vmps_2012supp}
Supplementary Material to Ref.~\onlinecite{guo_vmps_2012}.

\bibitem{suzuki}
M. Suzuki, Prog. Theor. Phys. {\bf 49}, 424, 1106, 1440 (1973).

\bibitem{BLTV05}
R.~Bulla, H.~Lee, N.~Tong, and M.~Vojta,
Phys. Rev. B {\bf 71}, 045122 (2005).


\bibitem{kv04}
M. Kircan and M. Vojta, \prb {\bf 69}, 174421 (2004).

\bibitem{ren_foot}
Technically, the RG treatment requires the introduction of a renormalized running coupling. Its initial value is determined by the bare coupling and suitable powers of the ultraviolet cutoff scale, with details depending on the RG scheme. In our case, the cutoff is $\w_c=1$, and we can identify the initial value of the running coupling with the bare coupling.

\bibitem{zitko_energy_2009}
R.~\v{Z}itko and T.~Pruschke,
Phys. Rev. B {\bf 79}, 085106 (2009).

\bibitem{weichselbaum_variational_2009}
A.~Weichselbaum, F.~Verstraete, U.~Schollw\"ock, J.~I. Cirac, and J.~von Delft,
Phys. Rev. B  {\bf 80}, 165117 (2009).

\bibitem{weichselbaum_QSpace_2012}
A.~Weichselbaum,
Ann. Phys.   {\bf 327}, 12 (2012).

\bibitem{saberi_matrix-product-state_2008}
H.~Saberi, A.~Weichselbaum, and J.~von Delft,
Phys. Rev. B {\bf 78}, 035124 (2008).

\bibitem{schollwock_2011}
U.~Schollw\"ock,
Ann. Phys. {\bf 326}, 96 (2010).

\bibitem{zhang_density_1998}
C.~Zhang, E.~Jeckelmann, and S.~R. White,
\prl  {\bf 80}, 2661 (1998).


\bibitem{weie_optimized_2000}
A.~Wei{\ss}e, H.~Fehske, G.~Wellein, and A.~R. Bishop,
Phys. Rev. B  {\bf 62}, R747 (2000).

\bibitem{nishiyama_numerical_1999}
Y.~Nishiyama,
Eur. Phys. J. B {\bf 12}, 547 (1999).

\bibitem{polyakov75} A. M. Polyakov, Phys. Lett. {\bf 59B}, 79 (1975).

\bibitem{goldenfeld} N. Goldenfeld, \textit{Lectures on Phase Transitions and the Renormalization Group}, Westview Press (1992).

\bibitem{tv_pc} T. Vojta, private communication.

\bibitem{pgkondo1}
D.~Withoff and E.~Fradkin, Phys. Rev. Lett. {\bf 64}, 1835 (1990).

\bibitem{pgkondo2}
M. Vojta and L. Fritz, Phys. Rev. B {\bf 70}, 094502 (2004);
L. Fritz and M. Vojta, Phys. Rev. B {\bf 70}, 214427 (2004).

\bibitem{edmft} Q.~Si, S.~Rabello, K.~Ingersent, and J.~L.~Smith, Nature
{\bf 413}, 804 (2001) and \prb {\bf 68}, 115103 (2003).

\bibitem{ssye}
S. Sachdev and J. Ye, Phys. Rev. Lett. {\bf 70}, 3339 (1993).




\end{thebibliography}
\end{document}